\documentclass[
reprint,
superscriptaddress,
amsmath, 
aps,
pre
]{revtex4-2}
\usepackage{xr} 
\usepackage{amssymb}
\usepackage{enumitem}
\usepackage{subcaption}
\usepackage{graphicx}

\begin{document}

\title{Simply Connected Topology in Perturbed Vortices and Field-Reversed Configurations}
\author{Taosif Ahsan}
\affiliation{Plasma Science and Fusion Center, Massachusetts Institute of Technology, Cambridge MA 02179}
\author{S.A. Cohen}
\affiliation{Princeton Plasma Physics Laboratory, Princeton University, Princeton, NJ, 08543 USA}
\author{A.H. Glasser}
\affiliation{Fusion Theory and Computation, Inc., 24062 Seatter Lane NE, Kingston, WA 98346 }

\preprint{APS/123-QED}

\date{\today}

\begin{abstract}
Zero-helicity vortices, such as Hill's vortex and field reversed configuration (FRCs), have long been assumed to be toroidal in topology. This paper proves this long-standing assumption false: even under arbitrarily small odd-parity (with respect to the $z$-axis of symmetry) transverse field perturbations, flux surfaces in the interior region become simply connected in topology. This work updates the previous topological categorization, open field lines and closed field lines, separated by an ellipsoid separatrix, to three new distinct categories: open field lines in the outermost region, closed field lines on torus flux surfaces in the intermediate region, and closed field lines on simply connected flux surfaces in the innermost region. While the closed and open field lines are still separated by a shifted ellipsoid outer separatrix, a new crescent-shaped inner separatrix appears that separates the torus and simply connected surfaces. The simply connected region is significant even for small perturbations. For example, in a spherical vortex, for a perturbation roughly $10\%$ of the background vortex field strength, the simply connected region is $\sim40\%$ of the total volume of the region with closed field lines. The analysis also proves the conjecture regarding the field line of these vortices retaining closedness under odd parity perturbation in the full three-dimensional context, completing the previous partial proofs in the two-dimensional context. In addition to analysis, preliminary numerical simulations of charged particle trajectories in magnetic confinement created from field reversed configuration under odd parity perturbation were conducted. Existence of crescent-like simply connected volumes was also observed in this context, even when gyro-radii were taken to be small compared to the system size. Given that FRCs are sustained by a rotating magnetic field with odd parity, these results motivate a revision of FRC-related fusion confinement physics. Furthermore, given the mathematical equivalence to Hill's vortex, this updates our topological understanding of fluid flow in a wide array of phenomena. 
\end{abstract}

\maketitle

\vspace{-9 pt}
\section{Introduction}
\vspace {-4 pt}
\label{sec: Introduction}
To improve the confinement of plasma, many magnetic-confinement fusion-reactor (MFE) designs have a toroidal topology, \textit{e.g.}, tokamaks, stellarators, reversed-field pinches, and spheromaks. In the spheromak category, there is a promising type of fusion confinement device, with spacecraft propulsion applications,  called the field-reversed configuration (FRC) \cite{bellan,steinhauer2011review, Paluszek2023} sustained by adding an odd-parity rotating magnetic field (RMF) to Hill's vortex-like background magnetic field structure \cite{CohMil, Cohen_2000, glasser}. The FRC-RMF system can be modeled as a perturbed zero-helicity structure, such as Soloviev equilibrium \cite{solovev} and, equivalently, Hill's vortex \cite{Hill, glasser}. These vortices have found further importance in broad and diverse fields such as accretion disks in astrophysics \cite{Barge1995, Lovelace1999, Meheut2012}, geophysical dynamics \cite{moffatt1991vorticity}, and even biological systems like jellyfish motion \cite{dabiri2010jellyfish, rosett2019vortex, Shadden2006}. Given an exactly equivalent mathematical structure, the work presented in this paper equally applies to a broad array of phenomena. 

This motivates a systematic analysis of zero-helicity vortex structures. Hill’s seminal 1894 paper, by solving the Euler equations, described a self-sustaining spherical vortex moving like a solid body within a fluid \cite{Hill}. Wan \cite{wan} proved that Hill’s vortex is an energy-maximizing system using the variational principle. Amick and Fraenkel \cite{amick} demonstrated the uniqueness of the solution. The long-held assumption is that the topology of these vortices is toroidal, which, as this paper will show, is not true in the presence of arbitrarily small perturbations.

The topological shape of the vortices is primarily characterized by the flux-surfaces, defined as level sets of flux functions. For axisymmetric vortices, the flux function forms a set of foliated tori. In fluid vortices, the field lines are equivalent to particle motion, and they thus trivially stick to the flux surfaces. In magnetic field vortices, when plasma particles remain near a particular foliation, the particle's gyroradius is the characteristic radial step size due to Coulomb collisions, which determines confinement. Particles with small gyro radii generally follow field lines, so the topology of flux-surfaces, which are a continuous and smooth collection of field lines, is vital in understanding particle confinement. In some toroidal devices, such as tokamaks, many particle trajectories significantly deviate from a flux-surface, forming drift surfaces, \textit{e.g.}, banana orbits \cite{Wesson}. Confinement is severely degraded because the characteristic step size increases to the banana width. Zero-helicity structures, such as FRCs, have mostly been treated as toroidal inside their bounding surface, the separatrix. The existence of non-toroidal flux-surfaces may have implications for plasma confinement and stability. 

Several studies have examined axisymmetric perturbed vortices \cite{keller, Moffatt1978, choi2020stability}, and numerical studies have been conducted on three-dimensional perturbations of Hill’s vortices \cite{orlandi2020instabilities,CohMil}. Axially non-symmetric perturbations have significant consequences for plasma confinement. Reference \cite{Alan} showed that an even-parity perturbation fully opens up closed field lines of toroidal plasma devices, predicting degraded plasma confinement, a discouraging result for the viability of FRC. (In this paper and in \cite{Alan, CohMil, Ahsan1}, parity of the perturbation is understood to be parity of transverse $x, y$ components with respect to the plane $z=0$.)

Nevertheless, additional discoveries significantly improved hope for plasma confinement using RMF-induced FRC systems. A preliminary version of the work in this paper was done in \cite{CohMil,Ahsan1}. Firstly, the resilience of field-line closure under odd-parity perturbation was observed in simulations done in \cite{CohMil}. However, conjectures based on simulation are not always valid, and rigorous analysis is preferable for understanding the overall picture.

\cite{Ahsan1} attempted to analytically prove the observations and partially succeeded in proving the closure by developing a mathematical object named modified flux function (MFF). \cite{Ahsan1} also found the exact range of perturbation magnitude that preserves closure. Given the topological nature of the conjecture, static analysis was sufficient for the study, which this paper will also assume. However, the analysis in \cite{Ahsan1} was done for a limited two-dimensional slab of the full three-dimensional system. The analytical justification behind the conjectures postulated in \cite{CohMil} in a real-life three-dimensional context remained elusive, significantly limiting the understanding of the system.

This paper resolves this limitation using tools from differential topology. It has succeeded in proving the validity of the conjectures made in \cite{CohMil} in a full three-dimensional context. Furthermore, this analysis elucidated a more coherent picture of topological categorization beyond just closed and open field lines. An unexpected result was proven: \textit{even under infinitesimal perturbations, simply connected flux-surfaces exist within zero-helicity vortices.} This was missed by the simulations, and a partial analysis was done in \cite{CohMil, Ahsan1} due to lack of access to the total picture. Given that the RMF-FRC system is equivalent to a wide class of perturbed Hill’s vortices, the conclusion should appear in many different contexts, as noted earlier. This paper also finds a physical interpretation of the modified flux function found in \cite{Ahsan1}.

The paper is also interested in real particle motion beyond somewhat abstract field lines. In the context of fluid mechanics, the field lines under study are equivalent to fluid velocity fields, and the conclusions apply directly to actual particle motion. However, in the case of Hill’s vortex created from magnetic field lines used in plasma confinement, the situation is more complicated. Generally, particle motion can be approximated to first order (in the ion-cyclotron radii/system size expansion) by helical motion around a magnetic field-line. But in the context of FRCs, the field goes to zero, and this approximation breaks down. Thus, we wanted to extend the study and investigate whether these simply connected topologies appear in the case of particle motion as well. The numerical calculations of particle motion in Soloviev equilibria also surprisingly demonstrated the simply connected topology. Even for very small gyro-radii, particle trajectories may, counterintuitively, strongly deviate from flux-surfaces, forming a volume with a simply connected boundary.

Thus, this paper has three primary advancements: it finds the full proof of the conjecture in \cite{CohMil}, discovers new simply connected topology in Hill’s vortices, and numerically shows the existence of simply connected topology in particle motion in perturbed field-reversed configurations.
\begin{figure*}[!ht]
\begin{subfigure}{0.3\textwidth}
    \includegraphics[width=\textwidth]{ 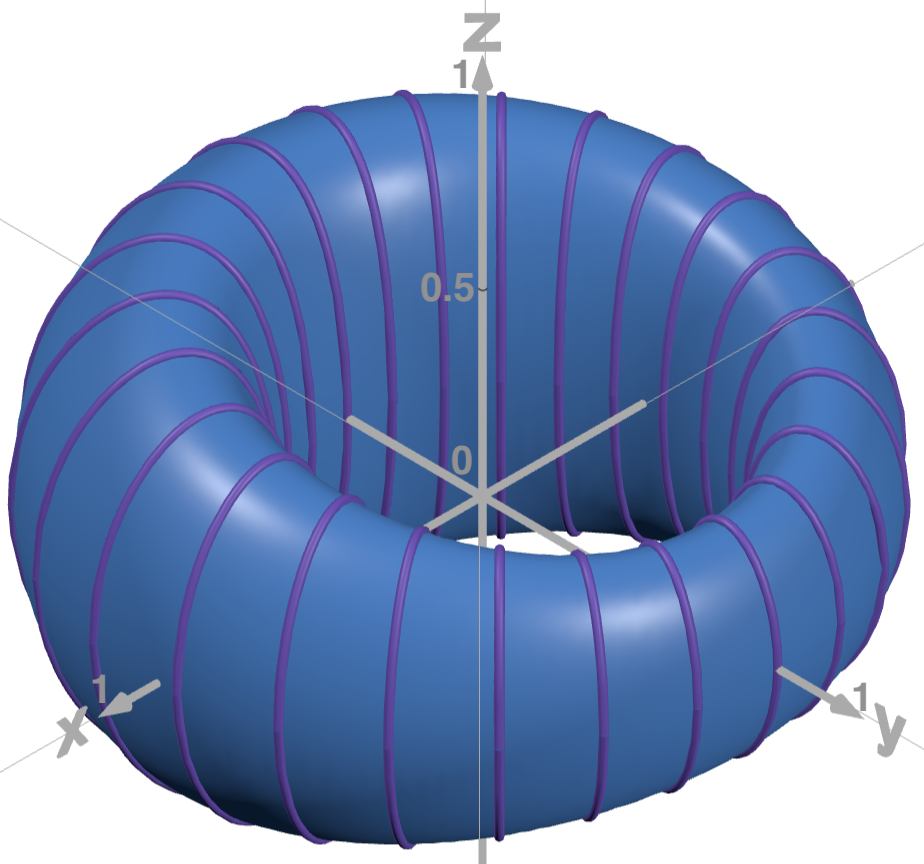}
    \caption{$\psi=0.165$ Wb. Torus.}
\end{subfigure}
~
\begin{subfigure}{0.3\textwidth}
    \includegraphics[width=\textwidth]{ 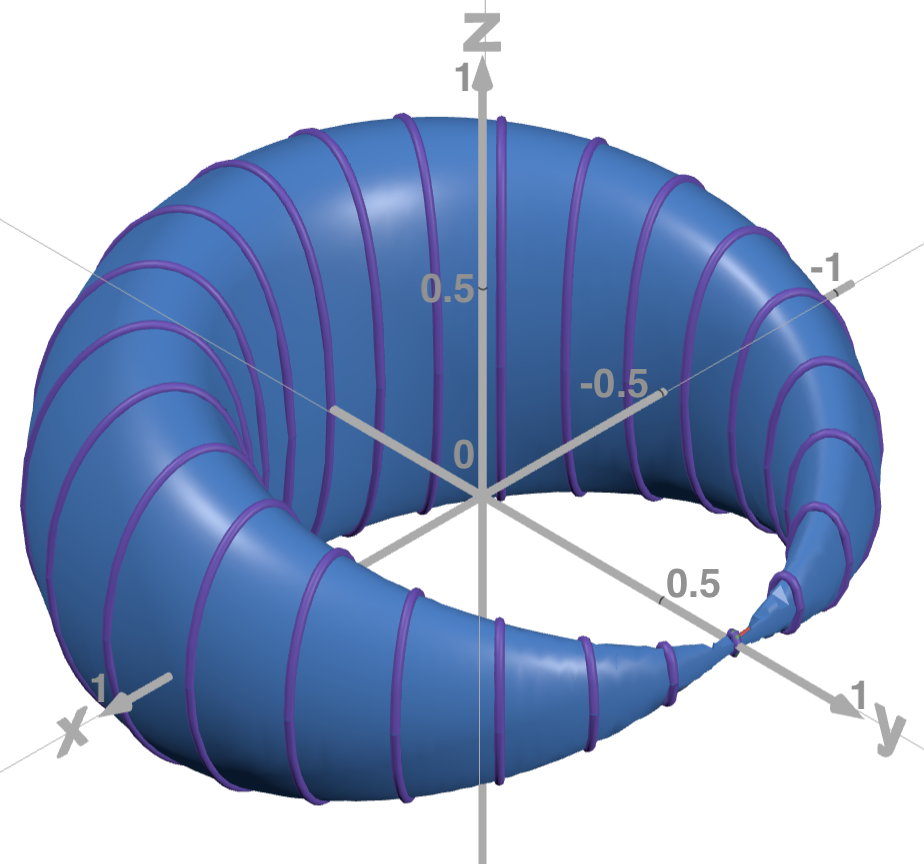}
    \caption{$\psi=0.195$ Wb. Transition.}
\end{subfigure}
~
\begin{subfigure}{0.3\textwidth}
    \includegraphics[width=\textwidth]{ 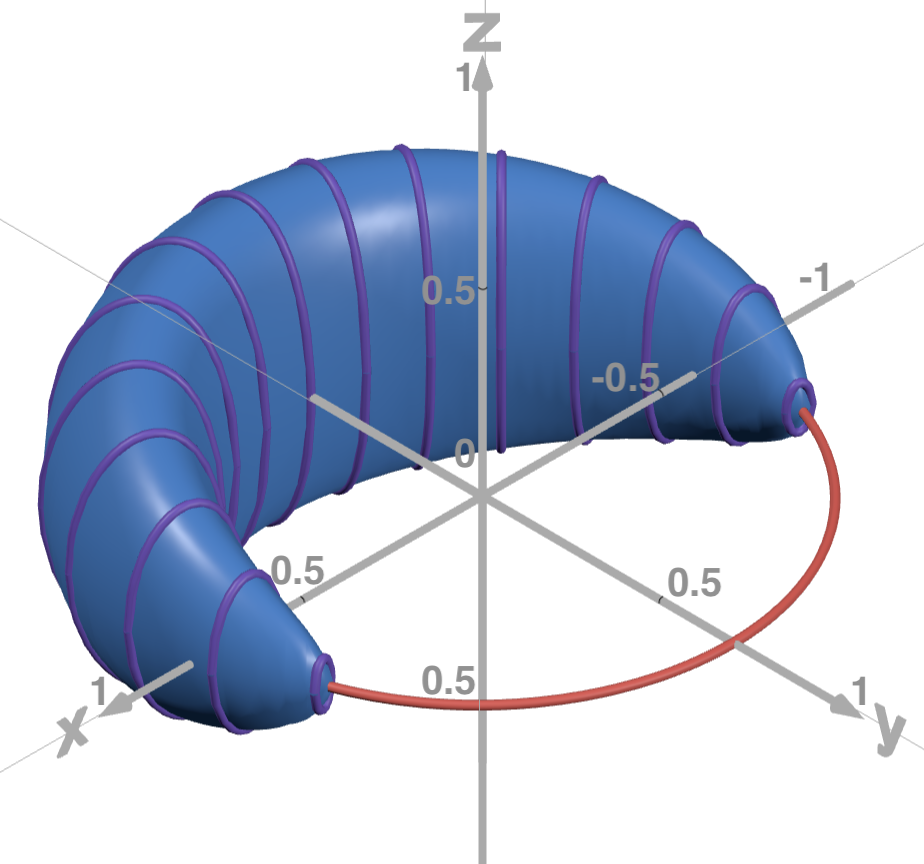}
    \caption{$\psi=0.23$ Wb. Simply connected.}
\end{subfigure}
    \caption{The flux-surfaces (blue) and field-lines (purple) transition from toroidal to simply connected as  $\psi$ increases. The red circle is the `O'-point null line on which the critical points $\textbf{r}_c$, defined by $\textbf{B}(\textbf{r}_c)=0$, lie. In (a), there is no intersection between the flux-surface and the red circle; in (b), there is one intersection; and in (c), there are two intersections.  Here, $\alpha = 0.2,\ k = 0.25\ \text{m}^{-1},\ B_0 = 2\ \text{T},\ r_s = z_s = 1\ \text{m}$.}
    \label{fig:fluxsurface}
\end{figure*}

\vspace {-4 pt}
 \section{Modeling the Perturbed Vortex with Modified Flux Function}
\vspace {-4 pt}
\label{sec: Modeling the Perturbed Vortex with Modified Flux Function}
In cylindrical coordinates, $(r,\phi,z)$, the vector field of an axisymmetric zero-helicity vortex, \textit{e.g.}, the magnetic field of an FRC, may be represented analytically by the Soloviev equilibrium or Hill’s vortex:
\begin{gather}
\textbf{B}_0 = B_0\left( \frac{r z}{z_s^2},\ 0, \  1-\frac{2r^2}{r_s^2}-\frac{z^2}{z_s^2}\right),
\label{eq: field-lines without perturbation}\\
\psi_{0} = \frac{B_0 r^2}{2}\ \left(1-\frac{r^2}{r_s^2}-\frac{z^2}{z_s^2}\right).
\end{gather}
$\textbf{B}$ can describe magnetic fields for plasma confinement or velocity fields in fluids, depending on context. $B_0$ is the vector field strength scaling. Without perturbation, the flux function $\psi_{0}$ and azimuthal angle between $r$ and cartesian $y$ axis, $\phi$, uniquely label field-lines. We perturb the vortex by $\textbf{B}=\textbf{B}_0+\delta\textbf{B}$,
\begin{equation}
\delta\textbf{B} = -\alpha B_0(kz\cos{\phi},-kz\sin{\phi},kr\cos{\phi}),
\label{eq: perturbation approx}
\end{equation}
where $\alpha>0$ is a free parameter controlling the perturbation’s field strength. $k>0$ is the axial wavenumber of the perturbation. The ratio of the field strength of a perturbation to the vortex field is $\sim\alpha k (r_s+z_s).$ This perturbation was first added to study the effect of long-wave odd-parity rotating magnetic field (RMF) on FRC. Despite a simple form, the perturbation model is rather general and includes any slowly spatially varying vacuum field perturbation that does not destroy closure. A detailed discussion of the perturbation model is added in appendix \ref{sec: Generality of the Perturbation Model}.

The model for the unperturbed vortex did not assume any drop in electric current/vorticity outside of the separatrix. This is mathematically valid but physically unsound; in a more realistic model, the vorticity drops to zero gradually in a thin layer. The topological analysis, however, remains physically valid because the relevant surfaces and points are all within the separatrix. However, under perturbation, the perturbed separatrix is slightly shifted outside, where the model can introduce some error in the thin crescent-like region where the perturbed and unperturbed separatrixes do not overlap. Fortunately, the error in field magnitude remains second order $\sim\alpha B_0\mathcal{O}(k^2r_s^2+k^2z_s^2)$ in that thin crescent-like region for reasonable assumptions on how quickly vorticity drops to $0$. This can be self-consistently ignored. See appendix \ref{sec: Validity of the Vortex Model} for a more detailed, rigorous discussion on this.

\begin{figure*}[t]
\begin{subfigure}{0.6\textwidth}
\includegraphics[width=\textwidth]{ 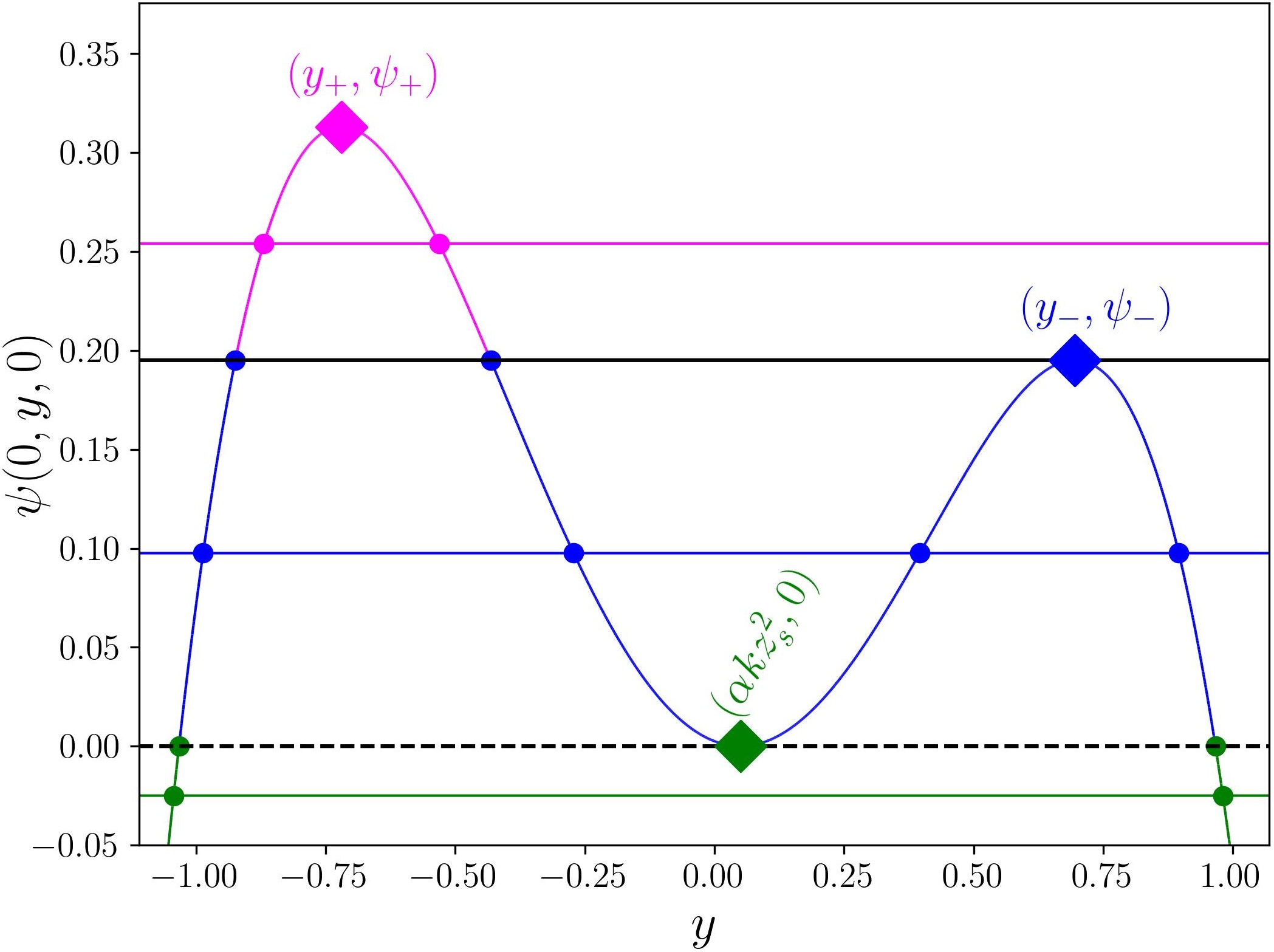}
\caption{$\psi(0,y,z)$ vs $y$}
\end{subfigure}

\begin{subfigure}{0.45\textwidth}
    \includegraphics[width=\textwidth]{ 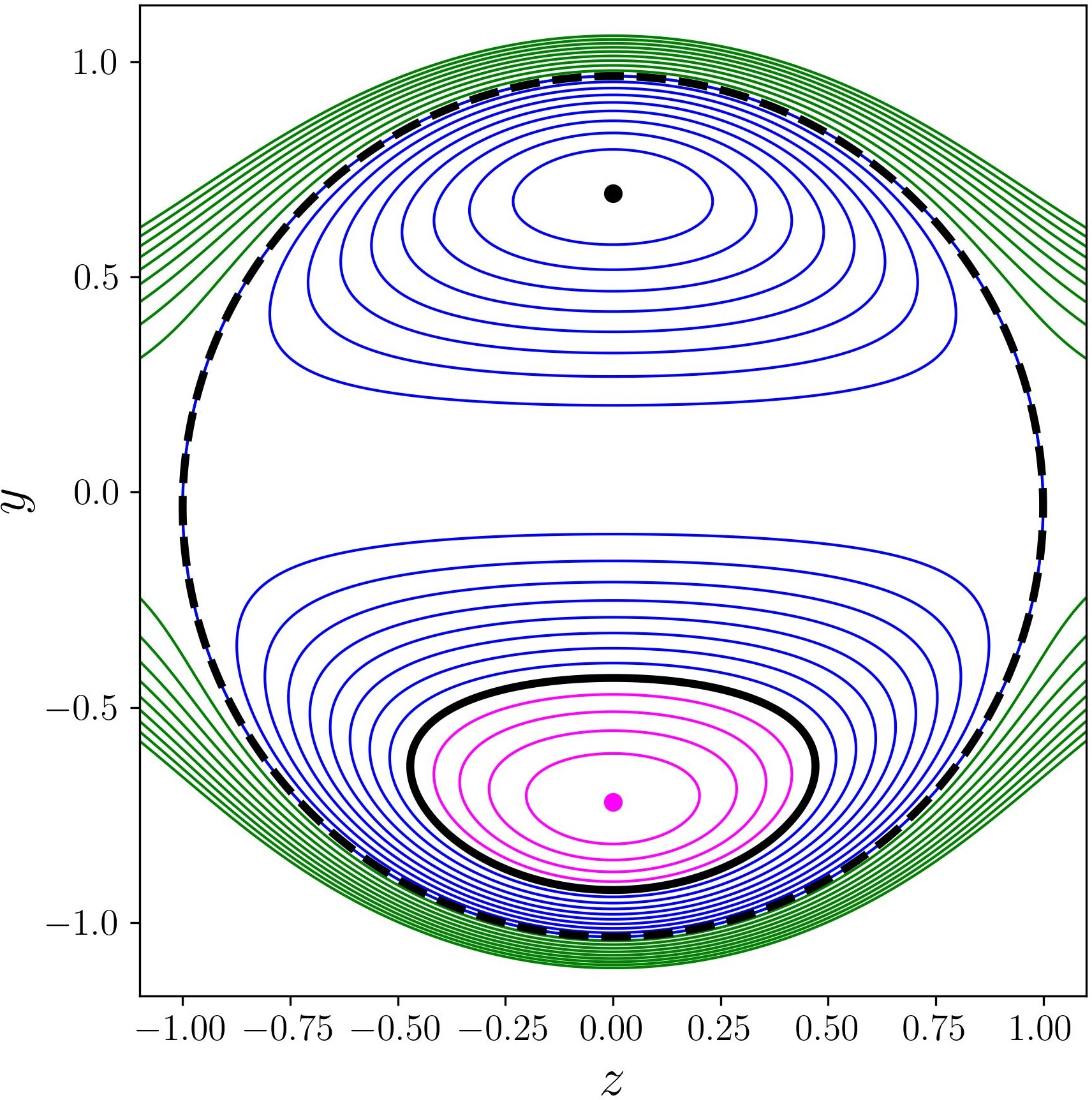}
    \caption{$\psi$ contours in $xy$ plane}
\end{subfigure}
~
\begin{subfigure}{0.45\textwidth}
    \includegraphics[width=\textwidth]{ 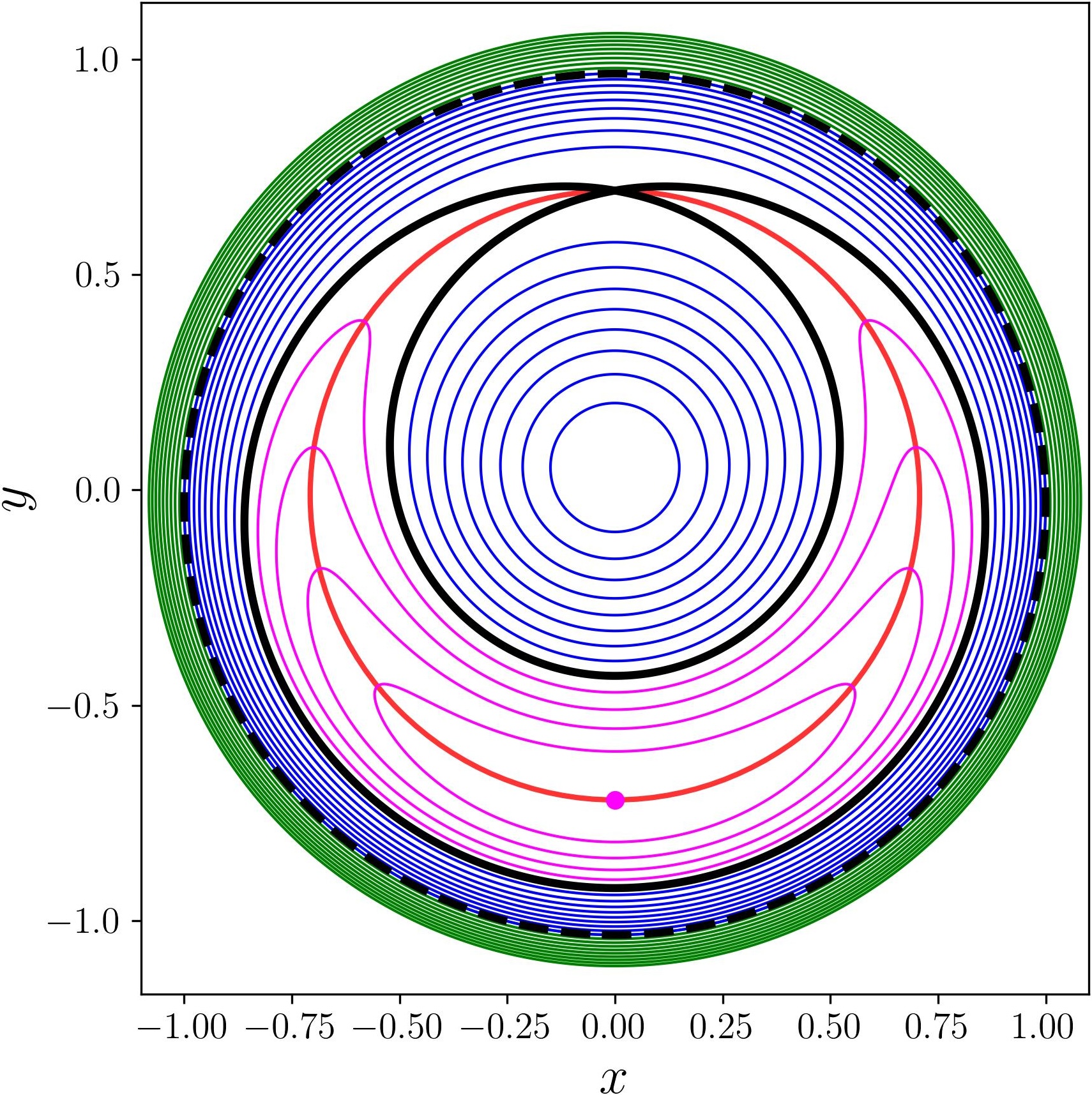}
    \caption{$\psi$ contours in $yz$ plane}
\end{subfigure}

\caption{(a) shows the $1$-dimensional $\psi(0,y,0)$ $y$ plot, with diamonds indicating the critical points. (b) and (c) are 2-dimensional plots of flux-surfaces intersecting with $yz$ and $xy$ planes, respectively. In all figures, green, blue, and magenta indicate open, torus, and simply connected topology, respectively. The dashed black and solid black lines indicate the outer and inner separatrix, respectively. (b) directly shows field-lines embedded on the $yz$ plane, while (c) can also be interpreted as the Poincaré map of field-lines on the $xy$ plane. The red circle is the $\textbf{B} = 0$ critical point circle in (c). Here, $\psi\in[-0.2,\psi_+=0.313]\ \text{Wb},\ \alpha = 0.2,\ k = 0.25\ \text{m}^{-1},\ B_0 = 2\ \text{T},\ r_s = z_s = 1\ \text{m}$.}
\label{fig:projection_surfaces}
\end{figure*}

With the addition of an odd-parity perturbation, the previous unique labeling fails because the system is no longer axisymmetric. A unique labeling for every field-line, a modified flux function (MFF), is still possible, as shown in \cite{Ahsan1}. A brief derivation of the modified flux function can be found if we translate the coordinate $\textbf{r}\rightarrow\textbf{r}-\alpha k z_s^2\ \hat{y}$. In this coordinate, the cylindrical radius and cylindrical coordinate azimuthal angle shift to
\begin{gather}
r\rightarrow\rho = \sqrt{(y-\alpha kz_s^2)^2+x^2},
\label{eq: rho}\\
\phi\rightarrow\varphi=\cot^{-1}\left(\frac{y-\alpha k z_s^2}{x}\right).
\label{eq: phi}
\end{gather}
As shown in appendix \ref{sec: Modified Flux Function}, this makes azimuthal component $B_\varphi = 0$ and gives us a set of Clebsch coordinate,
\begin{gather}
\textbf{B} = \nabla\psi \times\nabla\varphi
\label{eq: clebsch}
\end{gather}
The significance of the Clebsch pair $(\psi,\varphi)$ is that they are invariant along the flow and \textit{potentially} help uniquely label connected field-lines. One such set of field-lines is shown in Fig. \ref{fig:projection_surfaces}(c), where we have plotted level sets (flux-surfaces) of $\psi$ in $\varphi=0$ (or $yz$) plane. The only field lines this labeling scheme will miss are on the separator line (first identified in \cite{Ahsan1}),
\begin{gather}
    L \equiv\{\textbf{r}:\rho = 0\}
    \label{eq: separator line}
\end{gather}. 
Another physical interpretation of $\psi$ follows from
\begin{gather}
\psi = \frac{d\Phi}{d\varphi} =\int_0^\rho B_z\ \rho^\prime d\rho^\prime
\label{eq: flux definition}
\end{gather}

where $d\Phi$ is the total flux flowing through the infinitesimal arc situated between $[\varphi,\varphi+d\varphi]$ with radius $\rho$. The final physical interpretation of $\psi$ is given by magnetic vector potential $\textbf{A} = (\psi/\rho)\ \hat{\varphi}$, up to gauge. Hence, we see that, even though the modified flux function was first derived as a mathematical tool in \cite{Ahsan1}, it has multiple grounded physical interpretations. 

Using Eq. \eqref{eq: flux definition}, one can re-derive the form of $\psi$ found in \cite{Ahsan1}. In cartesian coordinates $(x,y,z)$, where $x=r\sin\phi,\ y=r\cos\phi,\ z=z$, $\psi$ can be written in the same form from \cite{Ahsan1},
\begin{gather}
\psi = \frac{B_0 r_s^2}{2}\cdot \textbf{u}^2(J-\textbf{v}^2)
\label{eq:psi}\\
\text{where,}\quad \textbf{u}=\left(\frac{x}{r_s},\ \frac{y-\alpha kz_s^2}{r_s},\ 0\right),
\label{eq: u}\\
\textbf{v}=\left(\frac{x}{r_s},\ \frac{y}{r_s}+\frac{\alpha kr_s}{3}\left(\frac{z_s^2}{r_s^2}+1\right),\ \frac{z}{z_s}\right),
\label{eq: v}\\
\text{and}\ J=1-\left(\frac{\alpha k r_s}{3} \right)^2\left(\frac{z_s^2}{r_s^2}+1\right)\left(\frac{2z_s^2}{r_s^2}-1\right).
\label{eq: J}
\end{gather}
See appendix \ref{sec: Modified Flux Function} for details of the calculations done to derive $\psi$ and its physical interpretation.

\vspace {-4 pt}
\section{Topological Classifications}
\label{sec: Topological Classifications}
\vspace {-4 pt}
An important topological parameter from \cite{Ahsan1} is
\begin{align}
\alpha_{c} = \frac{1}{kz_s\sqrt{1+2z_s^2/r_s^2}}.
\label{eq: critical strength}
\end{align}

Work in \cite{Ahsan1} shows that at $\alpha = \alpha_c$, field-lines in the $y$-$z$ plane undergo a phase transition and a new non-smooth separatrix forms. However, for $\alpha \geq \alpha_c$, flux-surfaces and field-lines with the same $\psi$ become disconnected. This ruins the possibility of unique labeling for connected lines and surfaces, along with introducing considerably more complicated topology. We therefore limit our work to $0 < \alpha < \alpha_c$, a reasonable assumption for small perturbations.
\vspace{-4pt}
\subsection{Observation}
\vspace {-4 pt}
We note the following observation. In Fig. \ref{fig:fluxsurface}, $\psi$ is increased while $\alpha<\alpha_c$ is kept fixed. We see that the flux-surfaces start from a torus-like shape in Fig. \ref{fig:fluxsurface}(a), and at a critical value of $\psi$, the torus appears to develop a sharp cusp (see Fig. \ref{fig:fluxsurface}(b)). Afterwards, the surface seems to become simply connected (see Fig. \ref{fig:fluxsurface}(c)). From the literature, we only expect a transition from open to closed flux-surfaces (or field-lines) at $\psi=0$. Here, however, we are observing a new and distinct transition between different kinds of closed flux-surfaces. We are interested in understanding the mechanism behind this transition and in numerically determining when it occurs.
\vspace{-4pt}
\subsection{Intuitive Understanding of the Mechanism}
\vspace{-4pt}
This observed change in topology is intimately related to the perturbation of the field and how it shifts the flux surfaces and critical points.
 
The flux surfaces are collections of field lines of $\textbf{B}$ tangent to them. Hence, the flux surface topology is determined by the nature of the critical points $\textbf{r}_c$, defined by $\textbf{B}(\textbf{r}_c)=0$, on these surfaces. There exists a circle of critical points in the system, shown as the red circle in Figs. \ref{fig:fluxsurface}(c) and \ref{fig:projection_surfaces}(c).

Without perturbation, the flux surfaces do not intersect the circle. Hence, they contain no critical points and have a toroidal topology. However, under perturbation, the flux surfaces shift, and so does the circle, but not by an equal amount. This brings the circle in contact with some interior flux surfaces. These flux surfaces now contain critical points and undergo a change in topology. Precisely what the topology changes into is stated and quantified in the next subsections.
\vspace{-4pt}
\subsection{Definitions}
\vspace {-4 pt}
We define the following:
\begin{gather}
\psi_\pm\equiv\psi(0,y_\pm,0)
,\ y_\pm\equiv\frac{-\alpha k r_s^2\mp r_s\sqrt{\alpha^2 k^2r_s^2+8}}{4}
\label{eq: psi_pm}\\
S_{\text{in}} \equiv \{\textbf{r}:\psi(\textbf{r})=\psi_{\text{in}}\to\psi_-,\ \psi_{\text{in}}>\psi_-\}
\label{eq: innerseparatrix}\\
S_{\text{out}} \equiv \{\textbf{r}: \textbf{v}(\textbf{r})^2=J\}
\label{eq: outerseparatrix}
\end{gather}
Inner and outer separatrix are respectively defined as $S_{\text{in}}$ and $S_{\text{out}}$. Here, $\textbf{v}(\textbf{r})$ and $J$ are given by Eqs. \eqref{eq: v} and \eqref{eq: J}. The physical definitions of $y = y_{\pm}$ are the two isolated maxima of $\psi(0,y,0)$, with $y = \alpha kz_s^2$ being the minima; see Fig. \ref{fig:projection_surfaces}(a). $\psi_\pm = \psi(0,y_\pm,0)$ can thus be physically understood as the two local maximum values. These maxima are shown clearly in Fig. \ref{fig:projection_surfaces}(a), where the diamonds indicate critical points. 

We define the following useful domain of angles intended for $\rho>0$ points,
\begin{gather}
\Theta(\psi) =\begin{cases}
    [0,2\pi)\ \text{if}\ \psi\leq\psi_-\\
    [\varphi_{\text{min}}(\psi),\ 2\pi-\varphi_{\text{min}}(\psi)]\ \text{if}\ \psi>\psi_-
\end{cases}\\
\varphi_{\text{min}}(\psi)=\cot^{-1}\left[h(\psi)/g(\psi)\right]\\
h(\psi)\equiv \frac{2 - \alpha^2 k^2 z_s^2 
+\sqrt{\left(\frac{\alpha^2}{\alpha_c^2} - 1 \right)^2 + \frac{24\psi}{B_0 r_s^2}}}{\alpha k (1 + 4 z_s^2/r_s^2)}\\
g(\psi)\equiv\frac{1}{\sqrt{2}}\sqrt{1-\alpha k r_s^2 h(\psi)-2h(\psi)^2}
\end{gather}
We also define the following useful domain of $z$ for points with $\rho = 0$,
\begin{gather}
    Z_1 = (-\infty, z_-),\ Z_2 =\{z_-\},\ Z_3 = (z_-, z_+),\nonumber\\ Z_4 = \{z_+\},\ \ Z_5=(z_+,\infty)
\end{gather}
The set $D$ is intended to be the domain of labeling for field lines,
\begin{gather}
D = \{(\psi,\varphi):\varphi\in\Theta(\psi),\ \psi \leq \psi_+]\}\sqcup\{i\star\}_{i=1}^5
\end{gather}
We also define the set of all field lines to be $\Gamma$.
\begin{gather}
    \Gamma \equiv \{l\subseteq \mathbb{R}^3:l \text{ is maximal integral curve of } \textbf{B}\}
\end{gather}

\vspace{-4pt}
\subsection{Visualization}
\vspace {-4 pt}
The intersection with the $yz$ plane is shown in Fig. \ref{fig:projection_surfaces}(b), which is equivalent to the field-lines themselves embedded on the $yz$ surface. The intersection of the $\psi$ flux-surfaces with the $xy$ plane is shown in Fig. \ref{fig:projection_surfaces}(c), which is equivalent to the Poincaré map of field-lines. In all figures, we use green for the $\psi\in(-\infty, 0)$ range, blue for the $\psi\in(0,\psi_-)$ range, and magenta for the $\psi\in(\psi_-,\psi_+]$ range. We also use a dashed black line for $\psi=0$ and a solid black line for $\psi =\psi_-$.

In Fig. \ref{fig:projection_surfaces}(a), as $\psi = \text{const.}$ is increased from negative values (green section) to $\psi = 0$, the number of intersections with $\psi(0,y,0)$ increases from $2$ to $3$. Afterwards, the blue $\psi=\text{const.}$ lines intersect $\psi(0,y,0)$ four times. This indicates a potential topological transition. Indeed, in Fig. \ref{fig:projection_surfaces}(b) we see that this is the boundary where the system transitions from open lines and surfaces (green) to closed lines and surfaces, hence it is named the outer separatrix $S_{\text{out}}$ and is shown as the dashed black line in Fig. \ref{fig:projection_surfaces}(a).

The second important event occurs when $\psi$ is increased from the blue region to $\psi_-$, where the number of intersections in Fig. \ref{fig:projection_surfaces}(a) again drops to $3$. Afterwards, in the magenta section, the number of intersections drops to $2$. This indicates another potential topological transition at $\psi = \psi_-$. However, we observe no topological difference between the lines on either side of the solid black line $\psi=\psi_-$ in Fig. \ref{fig:projection_surfaces}(b), and it is tempting to conclude that no topological transition occurs, as was concluded in \cite{Ahsan1}. However, we note that the solid black line level set has an extra isolated point on the upper side of the $y$ axis. Fig. \ref{fig:projection_surfaces}(c) is revealing: the magenta flux-surfaces inside the crescent-like solid black boundary are all connected curves, while the blue flux-surfaces outside all have two disconnected halves. This indicates that the flux-surfaces are transitioning from toroidal topology to simply connected topology. We therefore denote this newly observed transition surface as the \textit{inner separatrix} $S_{\text{in}}$.

The third important event occurs when $\psi$ reaches $\psi_+$, which is the reference value of $\psi$ at $(0,y_+,0)$. This is the global maximum value of $\psi$. Beyond this point, flux-surfaces cease to exist, as can be seen in Fig. \ref{fig:projection_surfaces}.
\begin{figure*}[!ht]
\centering
\begin{subfigure}{0.46\textwidth}
    \includegraphics[width=\textwidth]{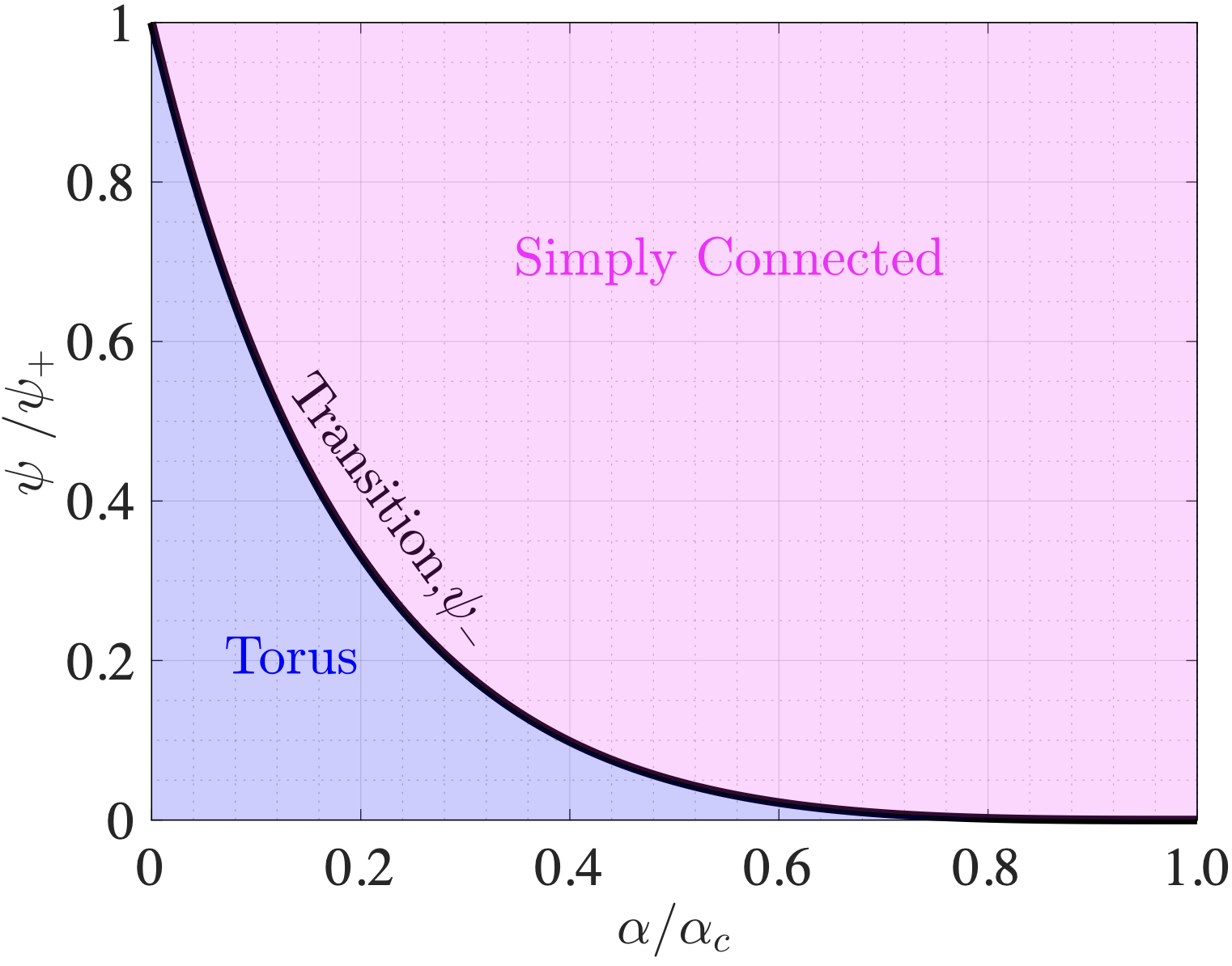}
    \caption{}
\end{subfigure}
~
\begin{subfigure}{0.5\textwidth}
    \includegraphics[width=\textwidth]{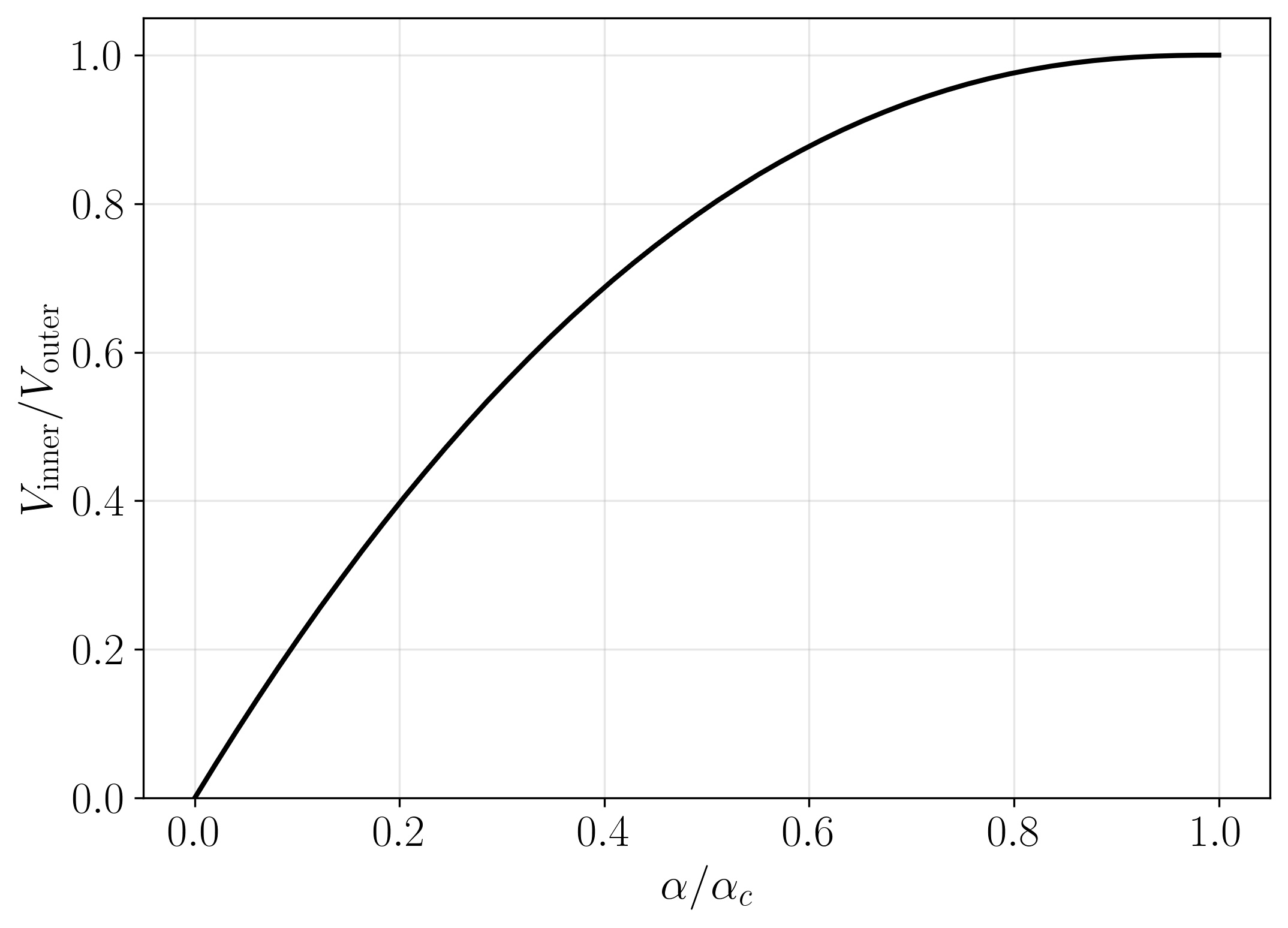}
    \caption{}
\end{subfigure}
\caption{(a) Categorization of compact flux-surfaces, $0<\psi<\psi_+$, (b) Volume ratio of inner separatrix and outer separatrix, for all $\alpha<\alpha_c$. Surfaces in the blue region, magenta region, and solid black line are toroidal, simply connected, and transitional. Here $r_s=z_s$.}
\label{fig:psi_vs_a}
\end{figure*}

\vspace{-4pt}
\subsection{Theorems}
\vspace{-4pt}
With these understandings, we can now state the following theorems to properly classify the topology of the flux-surfaces and field-lines.

For $0<\alpha<\alpha_c$,
\begin{enumerate}
\itemsep0em
\item Any field line is labeled by the bijection function $\gamma: D\to \Gamma$ given by,
\begin{gather}
    \gamma(\psi,\varphi) = \{\textbf{r}\notin  L:\psi(\textbf{r})=\psi,\ \varphi(\textbf{r})=\varphi\}\\
    \gamma(i\star) = \{\textbf{r}\in L: z\in Z_i\}
\end{gather}

\item For $\psi\in(-\infty,0)$, all field-lines are open lines and flux-surfaces are unbounded. They lie fully outside $S_{\text{out}}$.

\item For $\psi\in(0,\psi_-)$, all field-lines are closed loops and flux surfaces are homeomorphic to a torus. They lie strictly inside $S_{\text{out}}$ but outside $S_{\text{in}}$. 

\item For $\psi\in(\psi_-,\psi_+)$, all field-lines are closed loops and flux-surfaces are simply connected and homeomorphic to spheres. They lie strictly inside $S_{\text{in}}$. 
\end{enumerate}

\vspace{-4pt}
\subsection{Sketch of Proof}
\vspace{-4pt}
Proving the theorems requires a long and rigorous mathematical proof, given in Appendix \ref{sec: Proof of Theorems}. Here, we give a short sketch of the proof where we prove the following points successively. 
\begin{enumerate}
\itemsep0em
\item Points with $\psi > 0$ lie inside the outer separatrix, while those with $\psi < 0$ lie outside. Flux-surfaces and field-lines with $\psi > 0$ are compact, connected, and orientable.

\item Invariance of $(\psi,\varphi)$ on a connected field line and connectedness of field-lines with the same $(\psi,\varphi)$ implies an injective function from connected field-lines to $(\psi,\varphi)$ pairs.

\item These connected field-lines with $\psi < 0$ extend to infinity, implying that the field-lines and flux-surfaces are open.

\item $\psi_-$, the inner separatrix shown as a solid black line in Fig. \ref{fig:projection_surfaces}(c), is where flux-surfaces start intersecting critical points on the red circle.

\item Below $\psi_-$, flux-surfaces (shown as blue lines \ref{fig:projection_surfaces}(c)) lack critical points due to no intersection; above it (shown in magenta lines in \ref{fig:projection_surfaces}(c)), they contain two `O’-type critical points due to intersecting the critical red circle twice.

\item Consequently, from the \textit{Poincaré–Hopf theorem} \cite{poincare,hopf} and \textit{classification theorem}\cite{GuilleminPollack1974}, compact, connected, and orientable surfaces with $\psi>0$ inside the inner separatrix are simply connected (or spherical), while those outside are toroidal. (see Appendix \ref{sec: Proof of Theorems}.2)

\item The maximum possible $\psi$ is $\psi_+$, beyond which no flux-surfaces exist, and field lines are fully covered by the labeling in set $D$.
\item Compactness and connectedness of flux-surfaces imply that field-lines are closed loops, as they are intersections of flux-surfaces and $\varphi$ planes.
\end{enumerate}
\vspace{-4pt}

\begin{figure*}
\centering
\begin{subfigure}{0.40\textwidth}
   \includegraphics[width=\textwidth]{ 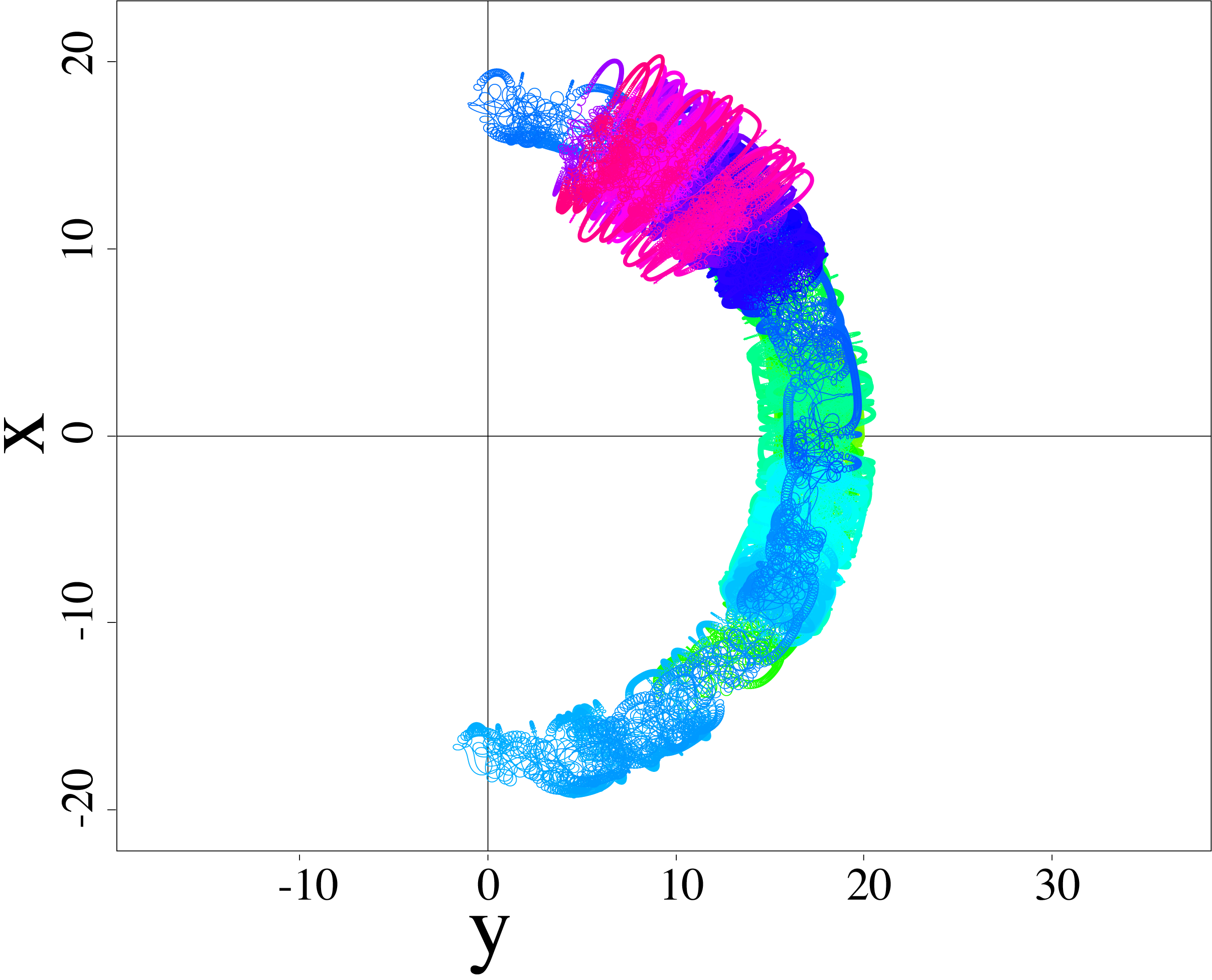}
   \caption{}
\end{subfigure}
~
\begin{subfigure}{0.40\textwidth}
   \includegraphics[width=\textwidth]{ 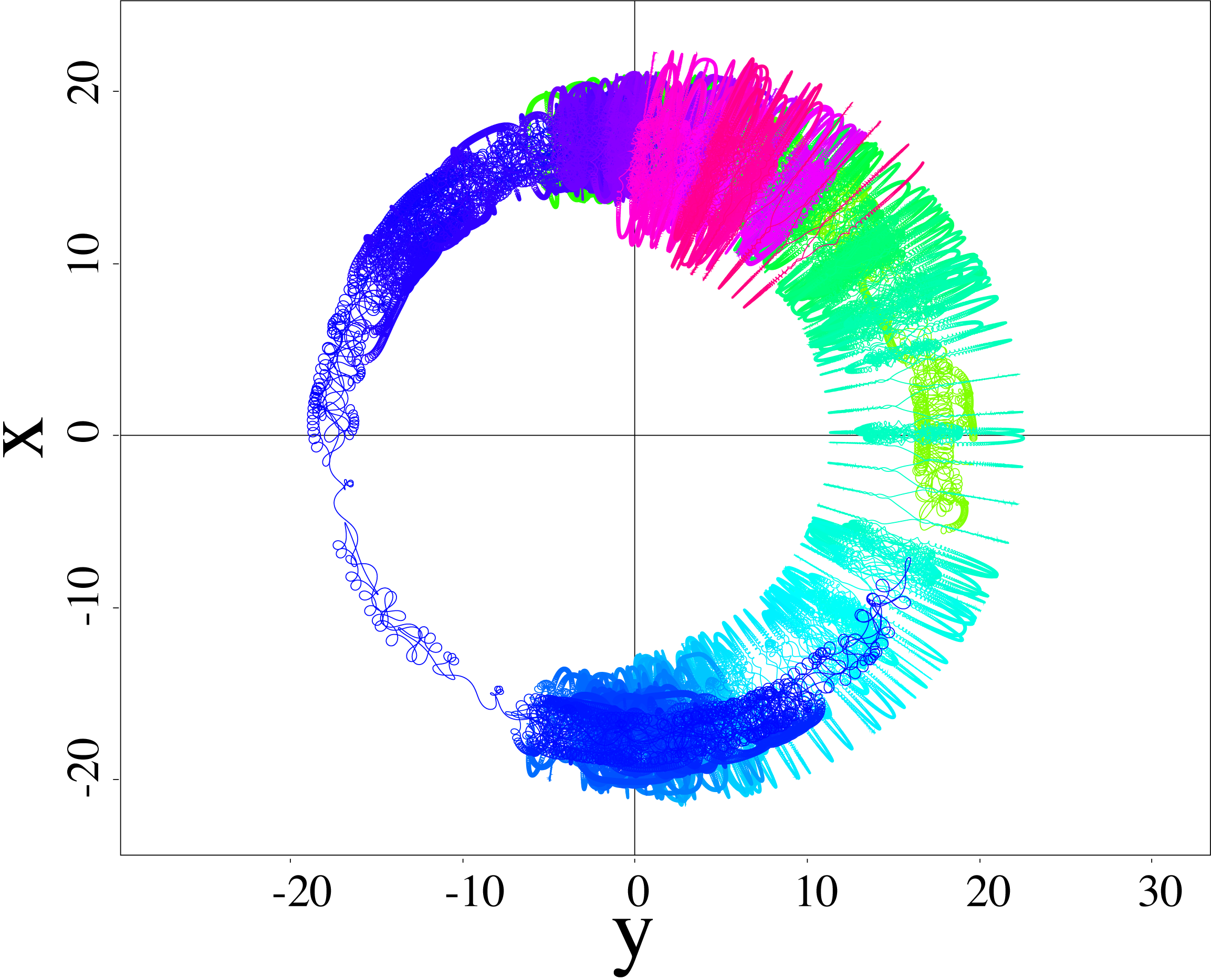}
   \caption{}
\end{subfigure}
~
\includegraphics[width=0.0405\textwidth]{ 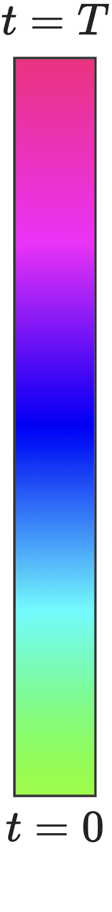}

\caption{Trajectories of electrons in perturbed FRC projected onto the $x$--$y$ plane. Unit in each axis is cm. Time is denoted by color. $\alpha = 0.1$, $I = 0.1$, $r_s = 25$ cm, $z_s = 75$ cm, $s\sim800$, $T=2\times10^5\tau_{ce}$ }
\label{fig: drift}
\end{figure*}

\subsection{Quantitative Results}
\vspace{-4pt}
$\alpha$ and $k$ can approach zero arbitrarily closely; thus, simply connected surfaces are guaranteed to exist. For a spherical vortex ($z_s = r_s$), with a small perturbation $\alpha = 0.2\alpha_c$ or $\alpha kr_s\approx 0.11$, the range of $\psi$ corresponding to compact, simply connected flux-surfaces remains $67\%$ of the total range of compact surfaces (see Fig \ref{fig:psi_vs_a} (a)). This increases to approximately $90\%$ at $\alpha = 0.4\alpha_c$ or $\alpha kr_s\approx 0.23$ and grows monotonically. These ratios were calculated using the closed-form expression given in Eqs. \eqref{eq: psi_pm}-\eqref{eq: outerseparatrix}.

The volume ratio of the simply connected region inside the inner separatrix and total compact surface region inside the outer separatrix is $40\%$ for $\alpha = 0.2\alpha_c$ or $\alpha kr_s\approx 0.11$, which increases to $69\%$ for $\alpha = 0.4\alpha_c$ or $\alpha kr_s\approx 0.23$. (see Fig. \ref{fig:psi_vs_a} (b)). $V_\text{inner}/V_{\text{outer}}$ was calculated using Monte Carlo integration, with $10^8$ samples per point, for $10^3$ uniformly spaced points of $\alpha \in (0, \alpha_c)$.

Even for perturbations of this small magnitude, these ratios are not negligible, indicating that the existence of simply connected surfaces is a robust feature of the Hill's vortices.
\vspace{-4pt}
\section{Particle Motion}
\vspace {-4 pt}
In a fluid vortex, the particle/fluid motion is the same as the field lines and is guaranteed to stick to the flux-surfaces discussed. Thus, particle motion has already been addressed in the previous section in the context of fluids. In the context of the $\textbf{B}$ representing FRC magnetic fields, particle motion deviates strongly from the field lines. The wide variety of motions can be attributed to nonlinearity, such as  $\mu$ non-conservation \cite{GlasseR_2002}. As representative cases, we examined, \textit{via} numerical simulation, whether electron motion exhibited simply connected patterns -- closed crescent-shaped surfaces -- in an FRC perturbed by static odd-parity RMF. 

A parameter commonly used to demarcate between fluid-like and kinetic FRC plasmas is \textit {s} $\equiv0.3r_s/r_g$, where $r_g$ is the particle gyro-radius at $r = r_s$ and  $z = 0$. (For $s > 10$ the FRC is generally considered fluid; for $s< 10$ it is generally considered kinetic.) This criterion is misleading because particles near the `O'-point null line or the `X'-point nulls experience a lower magnetic field, have larger $r_g$, and are in a region of greater field curvature, hence have a far lower local value \textit{s}. In the simulations described below,  \textit{s} $\sim 800$, but the local value of \textit{s}, \textit{$s_l$}, may be less than 1 in places along a particle's trajectory.

The simulations were performed with the  Hamiltonian code \textit{RMF} \cite{GlasserRSI}. Typical simulations computed the trajectories of electrons with $s=800$. Other relevant parameters were:  $r_s = 25$ cm, $z_s = 75$ cm; $B_0 = 50$ kG, $I = 0.1-0.5$ (where, $I\equiv kz_s/\pi$),  $\alpha = 0.01-0.1\ll \alpha_c = 0.73$, simulation duration, $T \sim2-4 \times 10^5 \tau_{ce} $ where $\tau_{ce} $ is the period of an electron's cyclotron motion at $r = z = 0$, and tolerances to numerical changes in the Hamiltonian ranged between $10^{-8}$ and $10^{-12}$. The electron's initial position was varied throughout the volume inside the (outer) separatrix. The initial velocity vector was similarly varied.  $\mu$ non-conservation decreases the average azimuthal velocity of these electrons below the thermal speed by a factor of 1000 and below the drift speed by a factor of 100. The relatively low azimuthal speed requires long-duration simulations to trace out the crescent shape.

Evident in the two cases shown in Fig. \ref{fig: drift}, the particle trajectories show crescent shapes. The orbits, despite the large \textit{s}, are not restricted to a surface. Crescent-shaped surfaces were seen up to $I\sim 0.5$, well beyond the validity of the long wave approximation.

For these long-duration simulations, the accumulated error in the Hamiltonian climbed to near $0.5\%$ for a tolerance of $10^{-8}$ but only to $10^{-4}\%$ for a tolerance of $10^{-12}$. For this range of tolerances, occasionally the crescent tips were connected, see Figure \ref{fig: drift}(a). Though that feature might be attributed to the  Hamiltonian method not preserving phase-space structure or to the Hairy Ball Theorem, we see tip-spanning trajectories only occur at $\mu$-non-conserving events near the crescent tips as B approaches 0. Approximately, $\mu$-non-conservation occurs when $s_l < 3$. 

That the trajectory occasionally makes azimuthal excursions close to but not precisely on the `O'-point null line --- see the blue and green trajectory segments in  Figure \ref{fig: drift}(b) ---   could infer that the drift surface has a toroidal geometry with the `O'-point line now serving as the ``major axis" of the new torus and the former major axis serving as the minor axis. A rigorous criterion for the shape of these drift surfaces is under study.

Yet another difference between the drift surface crescents and those of the modified flux function is that the former are often not symmetric about the perturbation. Additionally, it should be noted that crescents with clearly separated tips have only been seen in a small percentage of simulations.

Crescent-shaped trajectory surfaces are seen both when the applied perturbation is static or rotating \cite{glasser}. In the latter case, the interpretation is that charged particles are trapped in an azimuthal electric potential well. For the former, particles are trapped in a magnetic well. Depending on the direction of rotation of the perturbation, the crescents can overlay each other or be $\pi$ radians out of phase.
\vspace {-4 pt}
\section{Conclusions}
\label{sec: Conclusions}
\vspace {-4 pt}
The modification of the flux function used in the perturbed zero-helicity vortex and Soloviev equilibrium led to the discovery of simply connected flux-surfaces. The classification of flux-surface topologies and field-line closure was refined \textit{via} the calculation of an inner separatrix that separates simply connected and toroidal flux-surfaces. This challenges the prevailing notion that fluid vortices and fusion confinement structures exhibit purely toroidal topology and has implications for vortex and plasma dynamics. Observations about field line closure were rigorously proven in a full three-dimensional context. Simply connected topology is a robust feature in real-life stable vortices.

Numerical simulations of particle trajectories were conducted in the context of fusion confinement to calculate particle trajectories in a perturbed Soloviev equilibrium, where simply connected flux-surfaces were observed. The orbits often displayed crescent shapes, yet showed several marked differences from the flux-surfaces, \textit{e.g.}, gyro-radii of size comparable to the crescent diameter, asymmetric crescents, connections between well-separated crescent tips, and possibly toroidal shapes, with an inversion of the roles played by the major and minor axes. 
\vspace {-4 pt}
\section{Acknowledgments}
\vspace {-4 pt}
\label{sec: Acknowledgments}
The authors cordially thank Thanic Nur Samin for his advice. Support was provided, in part, by US DOE grants DE-FC02-99ER54512 and DE-AC02-09CH11466.
\vspace {-4 pt}
\section{Author Contributions}
\vspace {-4 pt}
\label{sec: Author Contributions}
TA: Conceptualization (lead); Formal analysis (lead); Methodology (lead); Visualization (lead); Writing – original draft (lead). SAC: Simulation (lead); Conceptualization (supporting); Methodology (supporting); Project administration (lead); Writing – review and editing (supporting). AHG: Methodology (supporting); Visualization (supporting).
\vspace {-4 pt}
\appendix
\section{Modified Flux Function}
\vspace{-4pt}
\label{sec: Modified Flux Function}
In Cartesian coordinates, the total perturbed magnetic field is,
\begin{gather}
    B_x = B_0 \frac{x z}{z_s^2},\\
    B_y = B_0 \left(\frac{y z}{z_s^2}-\alpha kz\right),\\  
    B_z =B_0\left(1-\frac{2(x^2+y^2)}{r_s^2}-\frac{z^2}{z_s^2} -\alpha k y\right),
\end{gather}
This is not symmetric with respect to $\phi$. However, switching to a new Cartesian coordinate with a translation $y' = y - \alpha k z_s^2$, gives us an axis-symmetric system. After simplification, we write in the new cylindrical coordinate $(\rho, \varphi, z)$,  
\begin{gather}
    B_\rho = B_0\frac{\rho z}{z_s^2},\\
    B_\varphi = 0,\\
    B_z =  B_0\left(1-\frac{\alpha^2}{\alpha_c^2}-\frac{2\rho^2+\alpha k (r_s^2+4z_s^2)\ \rho\cos\varphi}{r_s^2}-\frac{z^2}{z_s^2} \right)
\end{gather}
$\rho$ and $\varphi$ are given in Eqs. \eqref{eq: rho} and \eqref{eq: phi} and $\alpha_c$ is given in Eq. \eqref{eq: critical strength}. The system is still not axis-symmetric. However $B_\varphi = 0$. This allows us to define a modified flux function on $\rho > 0$ such that, 
\begin{gather}
    B_\rho = -\frac{1}{\rho} \frac{\partial \psi}{\partial z},\quad B_\varphi = 0,\quad B_z = \frac{1}{\rho}\frac{\partial \psi}
    {\partial \rho}
    \label{eq: shifted B psi}\\
    \implies \textbf{B} = \nabla\psi\times\nabla\varphi\\
    \implies \textbf{B}\cdot\nabla\psi = \textbf{B}\cdot\nabla\varphi = 0
    \label{eq: dot zero}
\end{gather}
This ensures that on a single field-line directed by $\textbf{B}$ field, $\psi$ and $\varphi$ are invariant for all points, and potentially can be used as a label. Every single field line $\in \mathbb{R}\backslash L$ is thus well defined by the two following level set equations,
\begin{gather}
    \psi(x,y,z) = \psi,\quad \varphi(x,y,z) = \varphi
\end{gather}
Thus, any field lines in $\rho>0$ can be labeled by $(\psi,\varphi)$. The edge case of $\rho = 0$ is handled in \ref{sec: Proof of Theorems}.3 \textit{lemma 0}. We will show in Appendix \ref{sec: Proof of Theorems}.3, \textit{lemma 3} that such labeling is unique for connected field lines by showing that all points with the same $(\psi,\varphi)$ are connected. 

There is also a physical meaning to $\psi$ beyond just a labeling scheme, 
\begin{gather}
    d\psi = \frac{\partial \psi}{\partial \rho}d\rho + \frac{\partial \psi}{\partial \varphi}d\varphi+\frac{\partial \psi}{\partial z}dz
\end{gather}
If we choose a path $c$ that keeps angular and $z$ coordinate fixed, while changing radial coordinate from $0$ to $\rho$, we get
\begin{gather}
    \psi = \psi(0,\varphi,z) + \int_c d\psi = \psi(0,\varphi,z) +\int_0^\rho\frac{\partial \psi}{\partial \rho^\prime}d\rho^\prime 
\end{gather}
We can, without any loss of generality, choose $\psi(0,\varphi,z) = 0$ and get,
\begin{gather}
    \psi = \int_0^\rho B_z\ \rho^\prime\ d\rho^\prime\\
    = \frac{B_0\rho^2}{2}\left(
1-\frac{\alpha^2}{\alpha_c^2}
-\frac{
\rho^2+\frac{2}{3}\alpha k (r_s^2+4z_s^2)\,\rho\cos\varphi
}{r_s^2}
-\frac{z^2}{z_s^2}
\right)
\end{gather}
Shifting back to the un-shifted Cartesian coordinate $(x,y,z)$ gives for the modified flux function outlined in Eqs. \eqref{eq:psi}. Given that $\psi \to 0$ as $\rho\to 0$ from all $\varphi$ direction, the construction of $\psi$ can be continuously extended to $\rho = 0$ line by assigning $\psi(\rho = 0) = 0$. However, no such continuous extension exists for $\varphi$ on the $\rho = 0$ line.

Total magnetic flux through an infinitesimal angular arc $[\varphi,\varphi+d\varphi]$ with radius $\rho$,  
\begin{gather}
   d\Phi = \int_\varphi^{\varphi+d\varphi}\int_0^\rho B_z\ \rho^\prime\ d\rho^\prime\ d\varphi^\prime
   \iff \frac{d\Phi}{d\varphi} = \psi
   \label{eq: physical definition}
\end{gather}
So, the physical interpretation of $\psi$ is the magnetic flux going through an infinitesimal angular arc $[\varphi,\varphi+d\varphi]$ with radius $\rho$. For the axis-symmetric case, this reduces to the usual total magnetic flux through a circular surface with radius $\rho = r$ normalized by $2\pi$.

There is another physical interpretation. We write the magnetic vector potential $\textbf{B} = \nabla\times\textbf{A}$ in $(\rho,\varphi,z)$ cylindrical coordinate,
\begin{gather}
B_\rho = \frac{1}{\rho}\frac{\partial A_z}{\partial \varphi}-\frac{\partial A_\varphi}{\partial z},\\
B_\varphi = \frac{\partial A_\rho}{\partial z}-\frac{\partial A_z}{\partial \rho},\\
B_z = \frac{1}{\rho}\frac{\partial (\rho A_\varphi)}{\partial \rho}-\frac{1}{\rho}\frac{\partial A_\rho}{\partial \varphi}.
\end{gather}
We can easily match this with Eq. \eqref{eq: shifted B psi} by choosing (up to gauge),
\begin{gather}
    A_\rho = 0,\quad \rho A_\phi =\psi,\quad A_z = 0
\end{gather}
So another physical interpretation of the modified flux function is,
\begin{gather}
    \textbf{A} = \frac{\psi}{\rho}\ \hat{\varphi}\quad \text{(up to gauge)}
\end{gather}
\vspace{-4pt}
\section{Proof of Theorems}
\vspace {-4 pt}
\label{sec: Proof of Theorems}
\subsection{Dimensionless Reduction of the Problem } 
\vspace{-4pt}
We make the system easier to analyze by making the following parameters dimensionless. 
\begin{gather}
    \tilde{x} \equiv  \frac{x}{r_s},\ \tilde{y}\equiv \frac{y}{r_s},\ \tilde{z} \equiv  \frac{z}{z_s},\
    \tilde{\psi} \equiv  \frac{2\psi}{B_0 r_s^2},\ \tilde{\textbf{B}}\equiv \frac{\textbf{B}}{B_0},\nonumber\\ \tilde{\alpha} \equiv  \alpha k r_s,\
    m \equiv  \frac{z_s^2}{r_s^2},\ \tilde{\alpha}_c \equiv  \alpha_c k r_s= \frac{1}{\sqrt{m(1+2m)}}
    \label{eq: replace}
\end{gather}
We will also denote the normalized Cartesian position vector to be,
\begin{gather}
    \tilde{\textbf{r}} \equiv \left(\tilde{x}, \tilde{y}, \tilde{z}\right)
\end{gather}
In terms of these parameters,
\begin{gather}
    \tilde{\textbf{B}}= \left(\frac{\tilde{x}\tilde{z}}{\sqrt{m}},\ \frac{(\tilde{y}-\tilde{\alpha}m)\tilde{z}}{\sqrt{m}},\ 1-2(\tilde{x}^2+\tilde{y}^2)-\tilde{z}^2-\tilde{\alpha}\tilde{y}\right)\\
    \tilde{\psi} = \textbf{u}^2(J-\textbf{v}^2)
    \label{eq: dimensionless MFF}
\end{gather}
The following parameters remain the same, so we will use the same symbols for our analysis.
\begin{gather}
    J=1-\tilde{\alpha}^2(m+1)(2m-1)/9\\
    \textbf{u}=(\tilde{x},\tilde{y}-\tilde{\alpha}m,0)\\
    \textbf{v}=\left(\tilde{x},\tilde{y}+\frac{\tilde{\alpha}(1+m)}{3},\tilde{z}\right)\\
    \varphi = \cot^{-1}\left(\frac{\tilde{y}-\tilde{\alpha}m}{\tilde{x}}\right) 
\end{gather}
In cylindrical co-ordinate $(u,\varphi,\tilde z)$ such that $\tilde x =  u\sin\varphi,\ \tilde y =  u\cos\varphi+\tilde\alpha m,\ \tilde z=\tilde z$. In terms of these co-ordinates, 
\begin{gather}
    \tilde\psi = u^2(I-u^2-a\ u\cos\varphi-\tilde z^2),\label{eq: reduced psi pol}\\
    \text{where, } I \equiv 1-\tilde\alpha^2/\tilde\alpha_c^2>0,\nonumber\\
    \text{and, } a \equiv  2\tilde\alpha(1+4m)/3>0\nonumber
\end{gather}

We will analyze everything for $\tilde{\alpha}<\tilde{\alpha}_c$ before the phase transition so that it will be an implicit assumption. Furthermore, we will not be using the symbol $\rho$ in the dimensionless case as $u=|\textbf{u}|$ already denotes the same parameter. 

\subsection{Necessary Constructions}
We first find and then define the following dimensionless sets (which we will show to be critical points), $\tilde{\textbf{B}}(\tilde{\textbf{r}}_c)=0$.
\begin{gather}
C_{\text{crit}} \equiv \{\tilde{\textbf{r}}:1-2(\tilde x^2+\tilde y^2)-\alpha\tilde  y =0,\ \tilde z=0\}
\label{eq: c_crit}\\
P_{\text{crit}}
\equiv\{(0,\tilde{\alpha} m, \tilde z_+),\ (0,\tilde{\alpha} m,\tilde  z_-)\}
\label{eq: p_crit}\\
\text{where,}\ \tilde z_\pm \equiv \pm \sqrt{1-\tilde{\alpha}^2/\tilde{\alpha}_c^2} 
\end{gather}
We write the dimensionless version of inside, outside, and the surface of the outer separatrix as,
\begin{gather}
E=\{\tilde{\textbf{r}}:J > \textbf{v}(\tilde{\textbf{r}})^2\}\\
E^\prime=\{\tilde{\textbf{r}}:J < \textbf{v}(\tilde{\textbf{r}})^2\}\\
\tilde{S}_{\text{out}}=\{\tilde{\textbf{r}}:J = \textbf{v}(\tilde{\textbf{r}})^2\}
\end{gather}
We wrote the dimensionless version of the separator line
\begin{gather}
L = (0,\ \tilde{\alpha}m,\ \tilde{z})
\end{gather}
We also define the flux surfaces $\tilde{S}$ and fibers $\gamma$,
\begin{gather}
    \tilde{S}(\tilde{\psi}_*) = \{\tilde{\textbf{r}}\in\mathbb{R}^3: \tilde{\psi}(\tilde{\textbf{r}})=\tilde{\psi}_*\}
    \label{eq: flux surface label}
\end{gather}
If $\gamma \cap L =\varnothing$,
\begin{gather}
    \gamma(\tilde{\psi}_*,\varphi_*) = \{\tilde{\textbf{r}}\in \mathbb{R}^3\backslash L:\tilde{\psi}(\tilde{\textbf{r}})=\tilde\psi_*,\ \varphi(\tilde{\textbf{r}})=\varphi_*\}
    \label{eq: field line label 1}
\end{gather}
If $\gamma \cap L \neq \varnothing$,
\begin{gather}
    \gamma(i\star) = \{\tilde{\textbf{r}}\in L: \tilde{z}\in Z_i\}
    \label{eq: field line label 2}\\
    Z_1 = (-\infty, \tilde z_-),\ Z_2 =\{\tilde z_-\},\ Z_3 = (\tilde z_-,\tilde z_+),\nonumber\\ Z_4 = \{\tilde z_+\},\ \ Z_5=(\tilde z_+,\infty)
\end{gather}
The way to differentiate the two types of $\gamma$ lines are labeling $\varphi_*\in[0,2\pi)$ vs. $\star$ labeled $i\star$ with integer $i$'s.

All $\tilde{S}(\tilde{\psi})$ are mutually disjoint because they are level sets of $\tilde{\psi}(\tilde{\textbf{r}})$. Similarly, all $\gamma(\tilde{\psi}_*,\varphi_*)$ are disjoint from one another because they are level sets of $(\tilde{\psi}(\tilde{\textbf{r}}),\varphi(\tilde{\textbf{r}}))$. Additionally $\gamma(i\star)\subseteq L$ and $\gamma(\tilde{\psi}_*,\varphi_*)\cap L =\varnothing$ making these categories of fibers disjoint. Given that $Z_i$'s are disjoint, all $\gamma(i\star)$ are also mutually disjoint. That is to say, all $\tilde{S}$ are mutually disjoint and all $\gamma$ are mutually disjoint. Thus, this construction can potentially qualify as a function from $(\tilde{\psi}_*,\chi)$ where $\chi=\varphi_*$ or $\chi = i\star$ to field lines. However, notice that we have not yet named these disjoint $\gamma$ as the field lines. One of the major goals of this appendix will be to prove that these disjoint fibers exhaustively label all field lines through an injective function. 

We note another change in convention, "closed/open" in the set theoretic sense and field line physics sense differ. From this point onward, we use the set-theoretic definition of "closed/open" and replace the physics-based "closed/open field line" with "bounded/unbounded" field line. We will revert to physics-based "closed/open" field lines when we finalize our proof.

\vspace{-4pt}
\subsection{Background Theorems}
\vspace{-4pt}
\label{app: poincarehopf}
\noindent\textbf{Poincare-Hopf Theorem }For a compact and differentiable manifold $M$ (such as flux-surfaces) with a continuous vector field $\textbf{A(\textbf{r}})$ on it, the indices of critical points $\textbf{r}_i$ (where $\textbf{A}(\textbf{r}_i)=\textbf{0}$) adds up to the Euler characteristics $\chi(M)$ of the manifold\cite{poincare, hopf}. 
\begin{gather}
    \chi(M)=\sum_{i}\text{index}_{\textbf{r}_i}(\textbf{A})
\end{gather}
Here, the index of the critical point means how field lines behave near it. If they are elliptically orbiting `O' points, they have an index of $1$; if they are hyperbolically orbiting `X' points, they have an index of $-1$, and so on. Spheres have $\chi(M)=2$, torus have $\chi(M)=0$.\\

\noindent\textbf{Classification Theorem }Every compact, connected, orientable surface without boundary is homeomorphic to a sphere $(g=0)$, a torus $(g=1)$, or a connected sum of $g$ tori for some integer $g \geq 0$, where $g$ is the genus number. The Euler characteristic is $\chi=2-2g$.

From this, $\chi(M) = 2\implies g = 0\implies$ sphere, and $\chi(M) = 0\implies g = 1\implies$ torus.\cite{GuilleminPollack1974}\\

\noindent\textbf{The Jordan-Brouwer Separation Theorem } Any compact, connected hypersurface $S$ in $\mathbb{R}^n$ will divide $\mathbb{R}^n$ into two connected regions; the interior $\text{int}(S)$ and the exterior $\text{ext}(S)$. $S,\ \text{int}(S),\ \text{ext}(S)$ are all disjoint to each other.\cite{GuilleminPollack1974}\\

\noindent\textbf{Regular Value Theorem} If $c$ is a regular value of $f$, meaning $\nabla f \neq 0$ for all points where $f=c$, then the level set $f^{-1}(c)$ is guaranteed to be a hypersurface.\cite{GuilleminPollack1974}
\vspace{-4pt}
\subsection{Proof of Lemmas}
\vspace{-4pt}
\noindent\textbf{Lemma 0 } Let $l$ be a field line with a point $\tilde{\textbf{r}}_1\in l$. If $\tilde{\textbf{r}}_1\in L$, then $l\subseteq L$. Otherwise, $l\subseteq\gamma(\tilde{\psi}(\tilde{\textbf{r}}_1),\varphi(\tilde{\textbf{r}}_1))$.

\noindent\textbf{Proof } Case $\tilde{\textbf{r}}_1\in L$: The field for all points $\in L$ has $\tilde{B}_{\tilde{x}}=\tilde{B}_{\tilde{y}}=0$. Thus, a field line that starts with the point $\tilde{\textbf{r}}_1\in L$ can not go outside the $L$ line, implying $l\subseteq L$.

Case $\tilde{\textbf{r}}_1\notin L$: If a point $\tilde{\textbf{r}}_2$ existed $\in l$ such that $\tilde{\textbf{r}}_2 \in L$ then from the earlier case $l\subseteq L$ implying $\tilde{\textbf{r}}_1\in L$ which is a contradiction. So $l$ has no point such that $u=0$. Given that $u>0$ everywhere on $l$, Eq. \eqref{eq: dot zero} readily applies. Let, $\tilde{\textbf{r}}_2$ be another point $\in l$.  

Integrating Eq. \eqref{eq: dot zero} from point $\tilde{\textbf{r}}_1$ to point $\tilde{\textbf{r}}_2$ on the field line $l$ gives,
\begin{gather}
\int_{\tilde{\textbf{r}}_1}^{\tilde{\textbf{r}}_2}  \frac{d\tilde{\textbf{r}}(s)}{ds} \cdot \nabla \tilde{\psi}(\tilde{\textbf{r}}(s))\ ds = 0
\implies \int_{\tilde{\textbf{r}}_1}^{\tilde{\textbf{r}}_2}  d \tilde{\psi}=0
\end{gather}
This means $\tilde{\psi}(\tilde{\textbf{r}}_2)= \tilde{\psi}(\tilde{\textbf{r}}_1)= $. An identical argument can show,  $\varphi(\tilde{\textbf{r}}_2)$. So, $\tilde{\textbf{r}}_2\in \gamma(\tilde{\psi}(\tilde{\textbf{r}}_1),\varphi(\tilde{\textbf{r}}_1)))$. Given that $\tilde{\textbf{r}}_2$ is a generic point in field line $l$, $l\subseteq \gamma(\tilde{\psi}(\tilde{\textbf{r}}_1),\varphi(\tilde{\textbf{r}}_1))$  $\square$\\

\noindent\textbf{Lemma 1 } $\tilde{\psi}(\tilde{\textbf{r}})> 0 \iff \tilde{\textbf{r}}\in E\backslash L$, $\tilde{\psi}(\tilde{\textbf{r}}) < 0 \iff \tilde{\textbf{r}}\in E^\prime\backslash L$ and $\tilde{\psi}(\tilde{\textbf{r}}) =0 \iff \tilde{\textbf{r}}\in \tilde{S}_{\text{out}}\cup L$. All points on $\tilde{S}(\tilde{\psi})$ has the vector field $\tilde{\textbf{B}}$ tangentially embedded on $\tilde{S}(\tilde{\psi})$

\noindent\textbf{Proof }If $\tilde{\textbf{r}}\in E\backslash L$ then ${\tilde{x}^2+(\tilde{y}-\tilde{\alpha}m)^2} >0$ and $J - \textbf{v}(\tilde{\textbf{r}})^2>0$ which implies $\tilde{\psi}>0$. If $\tilde{\psi}>0$, then either $\tilde{x}^2+(\tilde{y}-\tilde{\alpha}m)^2<0$ and $J - \textbf{v}(\tilde{\textbf{r}})^2<0$ or $\tilde{x}^2+(\tilde{y}-\tilde{\alpha}m)^2>0$ and $J - \textbf{v}(\tilde{\textbf{r}})^2>0$. The first is not possible, and the latter implies $\tilde{\textbf{r}}\in E$ and $\tilde{\textbf{r}}\notin L$.

If $\tilde{\textbf{r}}\in E^\prime\backslash L$ then ${\tilde{x}^2+(\tilde{y}-\tilde{\alpha}m)^2} >0$ and $J - \textbf{v}(\tilde{\textbf{r}})^2<0$ which implies $\tilde{\psi}<0$. If $\tilde{\psi}<0$, then either $\tilde{x}^2+(\tilde{y}-\tilde{\alpha}m)^2<0$ and $J - \textbf{v}(\tilde{\textbf{r}})^2>0$ or $\tilde{x}^2+(\tilde{y}-\tilde{\alpha}m)^2>0$ and $J - \textbf{v}(\tilde{\textbf{r}})^2<0$. The first is not possible, and the former implies $\tilde{\textbf{r}}\in E^\prime$ and $\tilde{\textbf{r}}\notin L$. 

If $\tilde{\textbf{r}}\in \tilde{S}_{\text{out}}\cup L$ then ${\tilde{x}^2+(\tilde{y}-\tilde{\alpha}m)^2} = 0$ or $J - \textbf{v}(\tilde{\textbf{r}})^2=0$ both of which imply $\tilde{\psi} = 0$. If $\tilde{\psi}=0$, then either $\tilde{x}^2+(\tilde{y}-\tilde{\alpha}m)^2 = 0$ or $J - \textbf{v}(\tilde{\textbf{r}})^2 = 0$ . This implies $\tilde{\textbf{r}}\in \tilde{S}_{\text{out}}\cup L$. 

For any point $\textbf{p}\in \tilde{S}(\tilde{\psi})$ with $u(\textbf{p})>0$ the associated field line $l\subseteq \gamma(\tilde{\psi},\varphi)\subseteq \tilde{S}(\tilde{\psi})$, see \textit{lemma 0}. Hence the vector field $\tilde{\textbf{B}}$ is tangent to $\tilde{S}(\tilde{\psi})$. If $u(\textbf{p})=0$ then $\tilde{\psi} = 0$ and $l\subseteq L\subseteq \tilde{S}(0)$ meaning the vector field $\tilde{\textbf{B}}$ is tangent to $\tilde{S}(0)$. $\square$\\

\noindent\textbf{Lemma 2 } All critical points $(\tilde{\textbf{B}}(\tilde{\textbf{r}}_c)=0)$ form the circle $C_{\text{crit}}$ or isolated critical point $P_{\text{crit}}$. They fulfill,
\begin{gather}
       C_{\text{crit}}\subseteq E\backslash L,\ P_{\text{crit}}\subseteq \tilde S_{\text{out}}\cap L
\end{gather}
Furthermore, critical points $\in C_{\text{crit}}$ are `O' types, while $\in  P_{\text{crit}}$ are `X' types.

\noindent\textbf{Proof } Critical points must fulfill $\tilde{\textbf{B}}(\tilde{\textbf{r}})=0$ condition.
\begin{gather}
    \frac{\tilde{x}\tilde{z}}{\sqrt{m}}=0,\
    \frac{(\tilde{y}-\tilde{\alpha}m)\tilde{z}}{\sqrt{m}}=0,\nonumber\\
    1-2(\tilde{x}^2+\tilde{y}^2)-\tilde{z}^2-\tilde{\alpha}\tilde{y}=0
\end{gather}

For the case $\tilde{z} = 0$, we can show that critical points form the following circle in the $\tilde z=0$ plane after simplification.
\begin{gather}
\tilde{x}^2+\left(\tilde{y}+\frac{\tilde{\alpha}}{4}\right)^2=R_1^2, 
\text{ where,  } R_1=\sqrt{\frac{1}{2}+\frac{\tilde{\alpha}^2}{16}}
    \label{eq: circle 1}
\end{gather}
The intersection of the $\tilde{S}_{\text{out}}$ sphere and $\tilde{z}=0$ plane is also a circle.
\begin{gather}
    \tilde{x}^2+\left(\tilde{y}+
    \frac{\tilde{\alpha}(1+m)}{3}\right)^2 = R_2^2\\
    R_2\equiv \sqrt{1-\frac{\tilde{\alpha}^2}{9}(m+1)(2m-1)}
    \label{eq: circle 2}
\end{gather}
The distance between the centers of the circles can be shown to be 
\begin{gather}
    d =\frac{\tilde{\alpha}(1+4m)}{12}<\frac{\tilde{\alpha}_c(1+4m)}{12}
\end{gather}
Radius of the circle in Eq. \eqref{eq: circle 1}
\begin{gather}
    R_1 < \sqrt{\frac{1}{2}+\frac{\tilde{\alpha}_c^2}{16}} 
    =\frac{(1+4m)\tilde{\alpha}_c}{4}\\
    \implies R_1+d<\frac{1+4m}{3\tilde{\alpha}_c}
\end{gather}
And of circle in Eq. \eqref{eq: circle 2}, 
\begin{gather}
    R_2>\sqrt{1-\frac{\tilde{\alpha}_c^2}{9}(m+1)(2m-1)}=\frac{1+4m}{3\tilde{\alpha}_c}\\
   \implies R_2>R_1+d   
\end{gather}

This means the circle from Eq. \eqref{eq: circle 1} is fully inside the circle Eq. \eqref{eq: circle 2}. So all $\tilde{\textbf{B}}(\tilde{\textbf{r}})=0$ points satisfy $\tilde{\textbf{r}}\in E$ for $\tilde{\alpha}<\tilde{\alpha}_c$. Thus $C_{\text{crit}}\subseteq E$. 

Now if the $\tilde{\textbf{r}}\in C_{\text{crit}}$ was also $\in L$ then,
\begin{gather}
    \tilde{x}=0,\ \tilde{y} = \tilde \alpha m,\ \tilde{z} = 0
\end{gather}
must fulfill,
\begin{gather}
    1-2\tilde\alpha^2 m^2-\tilde\alpha^2 m = 0
    \implies \tilde \alpha = \frac{1}{\sqrt{m(1+2m)}}
\end{gather}
which is not allowed. So, $C_{\text{crit}}\subseteq E\backslash L$.

Case $\tilde{z}\neq0$, $\implies \tilde{x}=0,\ \tilde{y}=\tilde{\alpha}m\ \implies \tilde{\textbf{r}}\in L$, and $ 1-2(\tilde{\alpha}m)^2-\tilde{\alpha}^2m-\tilde z^2=0$. This implies $\tilde z = \sqrt{1-\alpha^2/\alpha_c^2}$. After simplification, 
\begin{gather}
    \textbf{v}^2=(\tilde\alpha m+\tilde\alpha (1+m)/3)^2+1-m(1+2m)\tilde\alpha^2\nonumber\\
    =1-\frac{\alpha^2}{9}(m+1)(2m-1)=J
\end{gather}
Thus, $\tilde{\textbf{r}}\in \tilde S_{\text{out}}\implies P_{\text{crit}}\subseteq \tilde S_{\text{out}}\cap L$.

The field lines around these critical points lie on the flux-surface (as $\tilde{\textbf{B}}\cdot\nabla\tilde{{\psi}}=0$) and can thus be used to study the index of the critical points. The field-line is given by the equation, 
\begin{gather}
    \frac{d\tilde{\textbf{r}}}{ds}=\tilde{\textbf{B}}(\tilde{\textbf{r}})
\end{gather}
We do a first order expansion $|\delta\tilde{\textbf{r}}|\ll |\tilde{\textbf{r}}|$ in the neighborhood of the critical points $\tilde{\textbf{r}}$. Linearizing the equations and using the critical condition $\tilde{\textbf{B}}(\tilde{\textbf{r}})=0$ required gives us in cartesian coordinates, 
\begin{gather}
    \frac{d}{ds}(\delta\tilde{\textbf{r}})=\textbf{M}\ \delta\tilde{\textbf{r}},\nonumber\\
    \textbf{M} \equiv \left(\begin{matrix}
\tilde{z}/\sqrt{m} & 0 & \tilde x/\sqrt{m}\\
0 & \tilde{z}/\sqrt{m}  & (\tilde y-\tilde{\alpha}m)/\sqrt{m}\\
-4\tilde x & -4\tilde y-\tilde{\alpha} & -2\tilde{z}\\
\end{matrix}\right)
\end{gather}
The secular equation is $\det(\textbf{M}-\omega \textbf{I}) = 0$  with eigenvalue $\omega$ reduces to,
\begin{gather}
   \left(\frac{\tilde{z}}{\sqrt{m}}-\omega\right)\left[\left(\omega-\frac{\tilde{z}}{\sqrt{m}}\right)(\omega+2\tilde{z})+ g(\alpha,\tilde{x},\tilde{y})\right] = 0
    \label{eq: secular equation}\\
    \text{where,}\ g(\alpha,\tilde{x},\tilde{y})\equiv\frac{4\tilde{x}^2 + (4\tilde{y}+\tilde{\alpha})(\tilde{y}-\tilde{\alpha}m)}{\sqrt{m}}
\end{gather}

Case $\tilde{\textbf{r}}\in C_{\text{crit}}$: Replacing $\tilde x ^2$ using  Eq. \eqref{eq: circle 1}, and $\tilde{z}=0$, Eq. \eqref{eq: secular equation} further reduces to,
\begin{gather}
   -\omega\left(\omega^2+  \frac{f(\tilde{\alpha},\tilde y)}{\sqrt{m}}\right)=0,\\
     \text{where, }f(\tilde{\alpha},\tilde y)\equiv 2  - (1+4m) \tilde{\alpha} \tilde y - \tilde{\alpha}^2 m \nonumber
\end{gather}
We observe that $f$ is a strictly decreasing function of $\tilde y$. Furthermore, maximum possible $\tilde{y}$ in $C_{\text{crit}}$ circle is $R_1-\tilde \alpha/4$ or,
\begin{gather}
    \tilde{y}_- = \frac{-\tilde{\alpha}+\sqrt{\tilde{\alpha}^2+8}}{4}
\end{gather}
Thus, from inequality in Eq. \eqref{eq: solution range} we can conclude, 
\begin{gather}
    f(\tilde{\alpha},\tilde y_-)<f(\tilde{\alpha},\tilde{y})
    \label{eq: useful inequality 2}
\end{gather}
After some calculation, we find the expression of $f(\tilde{\alpha},\tilde{y}_-)$,
\begin{gather}
    f(\tilde{\alpha},\tilde{y}_-) = \frac{\sqrt{\tilde{\alpha}^2+8}}{4}\left(\sqrt{\tilde{\alpha}^2+8}-(1+4m)\tilde{\alpha}\right)
\end{gather}
From Eqs. \eqref{eq: useful inequality 1} and \eqref{eq: useful inequality 2}, we can conclude,
\begin{gather}
    0<f(\tilde{\alpha},\tilde{y}_-)<f(\tilde{\alpha},\tilde{y})
\end{gather}
One of the eigenvalues is $\omega = 0$. Given that $f(\tilde{\alpha},\tilde{y})>0$, the rest of the eigenvalues are imaginary; thus, the orbits around the critical points $\in C_{\text{crit}}$ are elliptical. This means $C_{\text{crit}}$ are `O' type critical points, but they are not isolated, which is expected given that they are on a circle. 

Case $\tilde{\textbf{r}}\in P_{\text{crit}}$: For these points $\tilde{x} = 0$, $\tilde{y} = \tilde{\alpha} m$ and $\tilde{z}=\tilde{z}_\pm$. Thus, the secular equation reduces to,
\begin{gather}
    \left(\omega-\tilde{z}/\sqrt{m}\right)^2(\omega+2\tilde{z}) =0
\end{gather}
Thus, all eigenvalues are real. Thus points $\in P_{\text{crit}}$ are `X' types. $\square$ \\

\noindent\textbf{Lemma 3 } All flux-surfaces $\tilde{S}(\tilde{\psi})$ are path connected surfaces and $\gamma$ fibers are path connected. 

\noindent\textbf{Proof } For $\tilde\psi = 0$, from \textit{lemma 1}, $\tilde{S}(\tilde{\psi}=0) = \tilde{S}_{\text{out}}\cup L$. $\tilde{S}_{\text{out}}$ is an ellipsoid which is connected and $L$ is an infinite connected line. Furthermore, $\tilde{S}_{\text{out}}\cap L\neq\varnothing$ because of \textit{lemma 2}. So $\tilde{S}(\tilde\psi=0)$ is a connected surface.

The $\gamma$ lines with $(\tilde{\psi},\varphi)$ labeling are cross section of the ellipsoid $\tilde{S}_{\text{out}}$ and $\varphi$ half planes. These lines are trivially connected. Furthermore, all $Z_i$'s are connected domains of $\mathbb{R}$, making $\gamma(i\star)$ also connected for all $i\star$. So any $\gamma$ on the level set of $\tilde\psi = 0$ is connected.

We now focus on $\psi\neq 0$ and $u>0$. In this region all $\gamma$ and are level sets of $(\tilde{\psi},\varphi)$. Furthermore, by definition,
\begin{gather}
    \tilde{S}(\tilde{\psi}) = \bigsqcup_{\varphi } \gamma(\tilde{\psi},\varphi)
    \label{eq: flux field union}
\end{gather}
We first show that all points in any fiber $\gamma(\tilde\psi,\varphi)$ are connected. Let us consider the section $\tilde z>0$ on a single $\gamma$-line $(\tilde\psi,\varphi)$. The $(\tilde\psi,\varphi)$ pair is a constant for a $\gamma$-line. This implies, 
\begin{gather}
    \tilde z^2 = I-u^2-au\cos\varphi-\frac{\tilde\psi}{u^2}
    \label{eq: z^2}\\\implies
    \tilde{z}\frac{d\tilde{z}}{du}=-2u-a\cos\varphi+\frac{2\tilde\psi}{u^3}
\end{gather}
We know $u>0$ because $\psi\neq0$ implies the point does not lie on the $L$ line, hence $u\neq0$. Furthermore, $z>0$ as per the assumption. So, the derivative exists for all points in the domain. Now, we prove that the domain of $\tilde z(u)$ is connected. 
We study the points where $z(u)=0$. This condition is satisfied when, 
\begin{gather}
    \tilde\psi = \Psi(u),\quad \Psi(u)\equiv u^2(I-u^2-au\cos\varphi)
\end{gather}
Have real solutions that happen when $\tilde\psi$ intersects $\Psi(u)$. We now set the derivative $d\Psi(u)/du$ to $0$ to find the critical points.
\begin{gather}
    u_0 = 0\quad u_\pm = \frac{1}{8} \left(-3 a\cos\varphi \pm \sqrt{9 a^2 \cos\varphi ^2+ 32 I}\right)
    \label{eq: critical}
\end{gather}
Given that $A>0$ and $I>0$, we have $9 A^2 \cos\varphi ^2+ 32 >(3A|\cos\varphi|)^2$. This gives us,
\begin{gather}
   u_+>\frac{3a}{8}(|\cos\varphi|-\cos\varphi)>0\\
\text{ and,}\  u_-<-\frac{3a}{8}(|\cos\varphi|+\cos\varphi)<0
\end{gather}
So, only one solution fulfills the $u>0$ condition. So, only a single critical point exists. It is not essential for our analysis if $u_c$ is a minima or a maxima; the conclusions remain the same. Given that only a single critical point exists,  $\tilde\psi$ intersects the $\Psi(u)$ vs $u$ plot in 0,1 or 2 points. Furthermore, $u$ can not have an unbounded domain as that makes $\tilde z(u)^2<0$ for large enough $u$.
\begin{enumerate}
    \item 0 points. For this case, the domain of $u$ is $(0,\infty)$, which is unbounded and hence impossible.
    \item 1 Point, let us denote the point $u_0$. The domain thus can be $(0,u_0)$ or $(u_0,\infty)$. $(u_0,\infty)$ is impossible because it is not bounded. So it must be $(0,u_0)$, which is simply connected.
    
    \item 2 points, let us denote the points $u_1$ and $u_2$ such $u_2>u_1$. Now, the domains can be $(u_1,u_2)$ or $(0,u_1)\cup(u_2,\infty)$.  $(0,u_1)\cup(u_2,\infty)$ is impossible because it is not bounded. So it must be $(u_1,u_2)$, which is simply connected.
\end{enumerate}
While ruling out $0$ points, we have also proven a strong claim: a $\gamma$-line must have at least 1 point where $z=0$.

Now, $\tilde z(u)$ is a continuous function of $u$ with a simply connected domain. So, the portion of the $\gamma$-line with $z>0$ is connected. Now, the system is mirror symmetric with respect to the $z=0$ plane. So, the portion of the $\gamma$-line with $\tilde z<0$ is also connected. Because of symmetry, the $z>0$ and $z<0$ portions of the $\gamma$-line $(\tilde\psi,\varphi)$ continuously limit to the same point on the $z=0$ plane. Thereby, any $\gamma$-line is connected.

Now, we show that all $\gamma$ lines on the same flux surface $\tilde\psi$ will have a continuous path connecting them. We first focus on the region $\tilde{x}\geq0$. 

If the flux surface only contains a single $\gamma$-line in this region, given that all points on a $\gamma$-line are connected, the $\tilde{x}\geq0$ portion of our flux surface is trivially connected. We now check the case where there exists more than a single $\gamma$ line on the $\tilde{x}\geq0$ portion of the flux surface. We pick two $\gamma$ lines on the $\tilde{\psi}$ surface, $\varphi_1$ and $\varphi_2$ such that $\varphi_1<\varphi_2$. We pick the point $a_1(u_*,\varphi_1,0)$ on $\tilde \psi$ surface and $\gamma$ line $\varphi_1$. Given that all $\gamma$ lines intersect the $z=0$ plane, such a point must exist. Here, $u_*$ is the radial value of the point, thus it satisfies,
\begin{gather}
    \tilde{\psi}=u_*^2(I-u_*^2-au_* \cos{\varphi_1})
\end{gather}
We define the following path,
\begin{gather}
    P_+(\varphi)=(u_*,\varphi,\tilde z_p(\varphi)),\ \text{ for all 
 }\varphi\in[\varphi_1,\varphi_2]\\ 
 \text{where, }
\tilde z_p(\varphi)=\sqrt{I-u_*^2-au_*\cos\varphi-\tilde\psi/u_*^2}
\label{eq: path}
\end{gather}
We can now deduce that, 
\begin{gather}
    z_p(\varphi)=\sqrt{I-u_*^2-au_*\cos\varphi_1-\tilde\psi/u_*^2} = 0\\
    \implies a_1(u_*,\varphi_1,0)\in P_+
\end{gather}
It is clear that path $P(\varphi)$ exists entirely on $\psi$ surface because after solving Eq. \eqref{eq: path} for $\tilde\psi$ we get, 
\begin{gather}
    \tilde\psi  = u_*^2(I-u_*^2-au_*\cos\varphi-\tilde z_p(\varphi)^2)
\end{gather}
Given that $\tilde x\geq0$, $\varphi_1, \varphi_2$  $\varphi$ must be within the range $[0,\pi]$. In this range, $\cos$ is a monotonically decreasing function. Thus, from the range of path $\varphi_1\leq\varphi\leq\varphi_2$ we have
\begin{gather}
    \cos{\varphi_1}\geq \cos{\varphi} \geq \cos{\varphi_2} 
\end{gather}
From Eq. \eqref{eq: path}, we see that $\tilde{z}_p^2$ is a strictly decreasing function of $\cos{\varphi}$. So, we finally conclude, 
\begin{gather}
    \tilde{z}_p(\varphi_1)^2\leq \tilde{z}_p(\varphi)^2\leq \tilde{z}_p(\varphi_2)^2\\
    \implies 0\leq\tilde{z}_p(\varphi)^2\leq \tilde{z}_p(\varphi_2)^2
    \label{eq: existence of line}
\end{gather}
This means all $\varphi\in[\varphi_1,\varphi_2]$ results in real existing points on $P_+$ and that also includes the endpoint that we denote as $a_2(u_*,\varphi_2,\tilde{z}_p(\varphi_2))$. Thus, there exists a continuous well defined path from $a_1$ on $\gamma$ line $(\tilde\psi,\varphi_1)$ to point $a_2$ on $\gamma$ line $(\tilde\psi,\varphi_1)$. Thus, any two $\gamma$ lines with $\tilde x>0$ have a continuous path connecting them. Given that all points in the $\gamma$ lines are also connected, this means all $\tilde x\geq0$ points on the flux surface are connected because of Eq. \eqref{eq: flux field union}. Given that the system is symmetric, all $\tilde x\leq0$ points on the flux surface are also connected.

We now determine whether these two halves are connected. We choose the point $b(u_*,\pi,\tilde z_p(\pi))$. We first need to show that the point actually exists. Given that $\varphi_1\leq\pi$, we can conclude as before that $z_p(\pi)^2\geq0$. This means the point is real. Secondly, solving for $\tilde{z}_p(\pi)$ we can write,
\begin{gather}
    \tilde{\psi} = u_*^2(I-u_*^2-au_*\cos(\pi)-\tilde z_p(\pi)^2)
    \label{eq: existence of pi}
\end{gather}

So the point $b(u_*,\pi,\tilde z_p(\pi))$ exists on the surface $\tilde{\psi}$. Hence, at least one point exists on the $\varphi=\pi$ plane, which is on the $\tilde{x} = 0$ plane. 

Given that $\tilde{x}=0$ does contain the point $b(u_*,\pi,\tilde z_p(\pi))$, all point in  $\tilde{x}\geq0$ and $\tilde{x}\leq0$ are connected to it from our earlier work. Hence, all points on any flux surface are connected. $\square$\\

\noindent\textbf{Lemma 4} All $\gamma$-fibers and flux-surfaces with $\tilde{\psi}<0$ are unbounded. $\gamma$-fibers exist for all $\varphi$.

\noindent\textbf{Proof} $\tilde{\psi}<0\implies u>0$, meaning $\gamma$-fibers fall under $(\tilde{\psi},\varphi)$ labeling. In terms of cylindrical coordinate $(u,\varphi)$, coordinate $\tilde z^2$ is given by Eq. \eqref{eq: z^2}. For $\tilde\psi<0$, the $-\tilde\psi/u^2$ term $\rightarrow\infty$ as $u\rightarrow0$ which gives us $\tilde{z}\rightarrow\pm\infty$. This means the $\gamma$-fiber is not bounded. The flux-surface is thus also unbounded. This also means that we can always find a sufficiently large $\tilde{z}$ for which $u$ exists regardless of $\varphi$. Hence $\gamma(\tilde{\psi},\varphi)$ is not empty for any $\varphi$. $\square$\\

\noindent\textbf{Lemma 5} Any flux-surfaces with $\tilde{\psi}>0$ are compact. 

\noindent\textbf{Proof }From \textit{lemma 1}, all points with $0<\tilde{\psi}$ exist inside the spherical finite region. So, any surface they create also exists in the bounded finite region. Thus, all flux-surfaces with $0<\tilde{\psi}$ are bounded. Furthermore, the level sets of any continuous function create closed sets. So the surfaces are compact. $\square$\\

\noindent\textbf{Lemma 6} All flux-surface associated with $\tilde{\psi}\notin\{0,\tilde\psi_-,\tilde\psi_+\}$ have no $\nabla\psi = 0$ points on them and are orientable. Here,
\begin{gather}
    \tilde{\psi}_\pm\equiv\tilde{\psi}(0,\tilde{y}_\pm,0) \text{\ and,\ } \tilde{y}_\pm\equiv\frac{-\tilde{\alpha}\mp\sqrt{\tilde{\alpha}^2+8}}{4}
    \label{eq: crit_definition}
\end{gather}

\noindent\textbf{Proof} Straightforward calculations can reduce $\nabla \psi = 0$ (written in rescaled cartesian coordinate) to,
\begin{gather}
    \tilde x (J-\textbf{u}^2-\textbf{v}^2)=0
    \label{eq: grad psi x}\\
    (\tilde y-\tilde\alpha m)(J-\textbf{u}^2-\textbf{v}^2)-\frac{\tilde \alpha(1+4m)\textbf{u}^2}{3}=0
    \label{eq: grad psi y}\\
   \textbf{u}^2\tilde z = 0
   \label{eq: grad psi z}
\end{gather}
Solving these gives us,
\begin{gather}
    (0,\tilde \alpha m,z)\quad\text{or,}\quad (0,\tilde y_\pm,0)
\end{gather}
But,
\begin{gather}
    \tilde \psi(0,\tilde\alpha m,z) = 0,\quad \tilde\psi(0,\tilde y_\pm,0)=\tilde \psi_\pm
\end{gather}
So these points belong to flux surfaces $\{0,\tilde \psi_-,\tilde \psi_+\}$. All other flux surfaces contain no point where $\nabla\psi = 0$. So, for any flux surface $\tilde \psi\notin\{0,\tilde \psi_-,\tilde \psi_+\}$, we can globally define a continuous, nowhere vanishing unit normal field  $\textbf{n}=\nabla\tilde \psi/|\nabla\tilde\psi|$. Thus all flux surface $\psi\notin\{0,\tilde\psi_-,\tilde\psi_+\}$ are orientable. $\square$\\

\noindent\textbf{Lemma 7} $\tilde{\psi}_+$ is the global maximum of $\tilde{\psi}$ and the following orderings are true:
\begin{gather}
    \tilde{y}_+<\tilde{\alpha}m<\tilde{y}_-,
    \label{eq: well ordered y}\\
    0<\tilde{\psi}_-<\tilde{\psi}_+
    \label{eq: well ordered}
\end{gather}

\noindent\textbf{Proof} Before proceeding further, we make a useful claim, 
\begin{gather}
    (1+4m)\tilde{\alpha}<\sqrt{\tilde{\alpha}^2+8}
    \label{eq: useful inequality 1}
\end{gather}
For if it was not true then,
\begin{gather}
\tilde{\alpha}^2+8\leq(1+4m)^2\tilde{\alpha}^2\\\implies 8 \leq((1+4m)^2-1)\tilde{\alpha}^2\\
\implies 1 \leq m(1+2m) \tilde{\alpha}^2\\
\implies \tilde{\alpha}_c\leq\tilde{\alpha}
\end{gather}
This violates $\tilde{\alpha}<\tilde{\alpha}_c$ proving Eq. \eqref{eq: useful inequality 1}. Now,
\begin{gather}
    \tilde{y}_--\tilde{\alpha}m = \frac{\sqrt{\tilde\alpha^2+8}-(1+4m)\tilde\alpha}{4}>0
    \label{eq: y_minus}
\end{gather}
For all $\tilde{\alpha}>0$,
\begin{gather}
    \tilde{y}_+=-\frac{\tilde{\alpha}+\sqrt{\tilde{\alpha}^2+8}}{4}<0<\tilde{\alpha}m
    \label{eq: y_plus}
\end{gather}
Eqs. \eqref{eq: y_minus} and \eqref{eq: y_plus} proves Eq. \eqref{eq: well ordered y}.

In Eq. \eqref{eq: reduced psi pol}, $\tilde\psi$ is a strictly decreasing function of $\tilde z^2$ and $\cos\varphi$. So setting $\tilde z = 0$ and $\cos\varphi=-1$ maximizes the function, which is on the $\tilde{y}$ axis. So, we attempt to find the maxima and minima points of  $\tilde\psi(0,\tilde y,0)$ on the $y$-axis. We can show after simplifying calculations, 
\begin{gather}
    \frac{\partial\tilde\psi(0,\tilde{y},0)}{\partial\tilde{y}}=2(\tilde y-\tilde \alpha m)(1-\tilde \alpha \tilde y-2\tilde y^2)
\end{gather}
Setting it to zero gives us our minima and maxima,
\begin{gather}
\tilde y = \tilde\alpha m\quad\text{and,}\quad \tilde y=\tilde y_\pm
\end{gather}
These critical points give us,
\begin{gather}
   \tilde  \psi(0, \tilde\alpha m,0) = 0,\quad\text{and,}\quad \tilde  \psi(0, \tilde y_\pm,0) = \tilde \psi_\pm
\end{gather}

Here, $y_\pm$ are reproduced as solutions of $1-\tilde \alpha \tilde y-2\tilde y^2=0$. So, $(0,\tilde y_\pm,0)$ lie on the critical circle $C_{\text{crit}}$ from Eq. \eqref{eq: c_crit}. From \textit{lemma 1} and $\textit{lemma 2}$ we can conclude $\tilde y_\pm\in E \backslash L \implies 0<\tilde{\psi}_\pm$.  

Furthermore, $\tilde{\psi}$ is maximized when $\cos(\varphi)=-1\implies \tilde{y}\leq\tilde{\alpha}m$ for global maxima. Hence the global maxima must be $(0,\tilde{y}_+,0)$ and $\tilde{\psi}_-<\tilde{\psi}_+$. This completes the proof for Eq. \eqref{eq: well ordered}. This also makes $\tilde\psi_+$ the global maximum of $\tilde \psi$. 
$\square$\\

\noindent\textbf{Lemma 8} Number of critical points of flux surface $\tilde{\psi}$ is given by 
\begin{gather}
n_{\text{crit}}(\tilde{\psi}) =
\begin{cases}
 2,\quad \psi\in(\psi_-,\psi_+)\cup\{0\}\\
 1,\quad \psi = \psi_\pm\\
 0,\quad \text{otherwise}
\end{cases}
\label{eq: intersect}
\end{gather}

\noindent\textbf{Proof}  Case, $0\leq\tilde\psi \leq \tilde \psi_+$: If $0<\tilde{\psi}<\psi_+$ contain any critical points, then it must intersect $C_{\text{crit}}$ as critical points on $P_{\text{crit}}$ are on $\tilde{\psi} = 0$ surface. The intersection point $(\tilde{x},\tilde{y},\tilde{z})$ must fulfill $\tilde{z}=0$ and, 
\begin{align}
    &\tilde{\psi}=(\tilde{x}^2+(\tilde{y}-\tilde{\alpha}m)^2)\nonumber\\
    &\quad\quad\times\left(J-\tilde{x}^2-\left(\tilde{y}+\frac{\tilde{\alpha}(m+1)}{3}\right)^2\right)\\
    &\quad\quad1-2(\tilde{x}^2+\tilde{y}^2)-\tilde{\alpha}\tilde{y}=0
    \label{eq: x intersect}
\end{align}
Where $\tilde y_\pm$ was defined in Eq. \eqref{eq: crit_definition}. Now replacing $\tilde{x}^2$ gives us after some simplifying steps, 
\begin{align}
    &\tilde{\psi} = 
\frac{1}{12}\left(1 + 2\tilde{\alpha}^2 m^2 - \tilde{\alpha} \tilde{y} (1 + 4  m) \right) \quad\nonumber\\
&\quad\quad\quad\times \left(3 - 2 \tilde{\alpha}^2 m (m + 1) - \tilde{\alpha} \tilde{y} (1 + 4 m) \right)
\label{eq: y intersect}
\end{align}
Solving the quadratic in Eq. \eqref{eq: y intersect} gives us two possibilities of $\tilde{y}=h_\pm$,
\begin{gather}
h_{\pm}(\tilde\psi)\equiv \frac{2 - \tilde{\alpha}^2 m 
\pm \sqrt{\left(\tilde{\alpha}^2/\tilde{\alpha}_c^2 - 1 \right)^2 + 12\tilde{\psi}}}{\tilde{\alpha}(1 + 4 m)}
\label{eq: h value}
\end{gather}
Replacing it in Eq. \eqref{eq: x intersect} gives us 4 possibilities of $\tilde{x}=g_\pm, -g_\pm$.
\begin{gather}
    g_\pm(\tilde{\psi})\equiv\frac{1}{\sqrt{2}}\sqrt{1-\tilde\alpha h_\pm(\tilde{\psi})-2h_\pm(\tilde\psi)^2}
    \label{eq: g value}
\end{gather}
So in total we have four possible pairs of $(\tilde{x},\tilde{y})\in\{(g_+,h_+),(-g_+,h_+),(g_-,h_-),(-g_-,h_-)\}$. For these to be valid solutions, they must fulfill, 
\begin{gather}
    0\leq g_\pm(\tilde{\psi})^2\\
    \implies 0\leq1-\tilde\alpha h_\pm(\tilde{\psi})-2h_\pm(\tilde\psi)^2\\  
    \implies \tilde{y}_+\leq h_\pm(\psi) \leq \tilde{y}_-
    \label{eq: solution range}
\end{gather}
We notice the following: $h_+$ is a strictly increasing function of $\tilde{\psi}$. Hence, the reverse must also be true, $\tilde{\psi}$ must be a strictly increasing function of $h_+$. This means that as we increase the $\tilde{y}$ component of the intersection points, the critical circle $C_{\text{crit}}$ intersects with different flux surfaces with increasing values of $\tilde{\psi}$. Thus,
\begin{gather}
\tilde{\psi}(0,\tilde{y}_+,0)\leq \tilde{\psi}(\pm g_+,h_+,0)\leq\tilde{\psi}(0,\tilde{y}_-,0)
\end{gather}
This leads to $\tilde{\psi}_+\leq \tilde{\psi}_-$ contradicting \textit{lemma 7}. Thus, the $(\pm g_+,h_+,0)$ points violate $g_\pm^2>0$ and are not real solutions.

We now focus on the $h_-$ branch of the solutions. Existence of a valid solution is equivalent to Eq. \eqref{eq: solution range}. $h_-$ is a strictly decreasing function of $\tilde{\psi}$. Thus, the inequality in Eq. \eqref{eq: solution range} is equivalent to, 
\begin{gather}
    \tilde{\psi}(0,\tilde{y}_+,0) \geq \ \tilde{\psi}(\pm g_-,\tilde{h}_-,0) \geq \tilde{\psi}(0,\tilde{y}_-,0)
\end{gather}
$(\pm g_-,\tilde{h}_-,0)$ are points on $\tilde{\psi}$ surface. $\tilde{\psi}_-\leq \psi\leq \psi_+$ is equivalent to flux surface $\tilde{\psi}$ having the following critical points,
\begin{gather}
(-g_-(\tilde{\psi}),h_-(\tilde{\psi})),\quad (g_-(\tilde{\psi}),h_-(\tilde{\psi}))
\end{gather}
However, it is possible for the intersection points to coalesce into a single point if $g_-(\tilde{\psi})=0$. This happens for the equality cases of Eq. \eqref{eq: solution range}. The equality cases are $h_-(\tilde{\psi})=\tilde{y}_\pm$. For these single critical points $\tilde{\psi}=\tilde{\psi}(0,\tilde{y}_\pm,0)=\tilde{\psi}_\pm$. Thus, 
\begin{gather}
    \tilde{\psi}_-<\tilde{\psi}<\tilde{\psi}_+ \iff n_{\text{crit}}=2 \text{, all from}\ C_{\text{crit}}
    \label{eq: equiv} \\
    \tilde{\psi}=\tilde{\psi}_\pm \iff n_{\text{crit}}=1 \text{, all from}\ C_{\text{crit}}
    \label{eq: equiv 1} 
\end{gather}
Furthermore, $0<\tilde{\psi}<\tilde{\psi}_-$ is disjoint from $\tilde{\psi}_-\leq \tilde{\psi}\leq \psi_+$ and thus they contain no critical points.
\begin{gather}
    0<\tilde{\psi}<\tilde{\psi}_- \iff n_{\text{crit}}=0
    \label{eq: equiv 2} 
\end{gather}

Case $\tilde\psi \leq 0$: From \textit{lemma 2}, the two critical points $\in P_{\text{crit}}\subseteq \tilde{S}_{\text{out}}$. From \textit{lemma 1}, they have $\tilde{\psi} = 0$. The other critical critical points are $C_{\text{crit}}\subseteq E\backslash L$, and thus must have $0<\tilde{\psi}$ because of \textit{lemma 1}. Thus, there are only two critical points on $\tilde{\psi} = 0$, and there are no critical points for $\tilde{\psi}<0$. 
\begin{gather}
    \tilde{\psi} < 0 \iff n_{\text{crit}}=0 
    \label{eq: equiv 3} \\
    \tilde{\psi} = 0 \iff n_{\text{crit}}=2 \text{, all from}\ P_{\text{crit}}
    \label{eq: equiv 4}
\end{gather}
Summarizing Eqs. \eqref{eq: equiv}-\eqref{eq: equiv 4} gives us Eq. \eqref{eq: intersect}.  $\square$\\

\noindent\textbf{Lemma 9} Flux-surface $\tilde{S}(\tilde{\psi})$ in the range $\tilde{\psi}\in(0,\tilde{\psi}_-)$ and $\tilde{\psi}\in(\tilde{\psi}_-,\tilde{\psi}_+)$ are respectively homeomorphic to torus and sphere. 

\noindent\textbf{Proof} Flux surfaces $\tilde{S}(\tilde{\psi})$ in the range $(0,\tilde{\psi}_-)$ and $(\tilde{\psi}_-,\tilde{\psi}_+)$ are smooth (see \textit{lemma 8} and regular value theorem) and compact (see \textit{lemma 5}). $\tilde{\textbf{B}}$ is smooth in $\mathbb{R}^3$ and tangent to $\tilde{S}(\tilde{\psi})$ (\textit{lemma 1}). Thus, Poincaré-Hopf theorem applies in this condition.

For $\tilde{\psi}\in(0,\tilde{\psi}_-)$, the total index must sum to $\chi(M)=0$ because there are no critical points (see \textit{lemma 8}). For $\tilde{\psi}\in(\tilde{\psi}_-,\tilde{\psi}_+)$, the total index must sum to $\chi(M)=2$ because there are two `O' type critical points (see \textit{lemma 2} and \textit{lemma 8}).

\textit{Lemma 3} tells us that the surfaces are connected. \textit{lemma 6} tells us that the surfaces in the range $(0,\psi_-)\cup(\psi_-,\psi_+)$ are orientable.

The only compact, connected, and orientable topology in $\mathbb{R}^3$ with $\chi(M)=0$ and $\chi(M) = 2$ are homeomorphic to respectively torus and sphere (see classification theorem\cite{GuilleminPollack1974}). This proves our theorem. $\square$\\

\noindent\textbf{Lemma 10} For all $0<\tilde\psi_1<\tilde\psi_2$, $\tilde S(\tilde \psi_2)$ is inside the interior of $\tilde S(\tilde\psi_1)$, assuming they are both hypersurfaces.

\noindent\textbf{Proof} We know from \textit{lemma 3} and \textit{lemma 5} that the flux surface $\tilde{S}(\tilde{\psi}_1)$ is closed, and connected. Given that by assumption it is a hypersurface, thus according to the Jordan-Brouwer Separation Theorem\cite{GuilleminPollack1974}, it separates $\mathbb{R}^3$ into two open connected components, bounded interior $\text{int}(\tilde\psi_1)$ and unbounded exterior $\text{ext}(\tilde\psi_1)$. Furthermore, $\tilde{S}(\tilde{\psi}_1)$,  $\text{int}(\tilde\psi_1)$ and  $\text{ext}(\tilde\psi_1)$ are all disjoint. 

Given \textit{lemma 1}, $\tilde \psi > 0$ is the ellipsoid $E$ (minus the line $L$) which is bounded in $\mathbb{R}^3$. Hence for large enough $|\textbf{q}|$, $\tilde \psi(\textbf{q})\leq 0$. $\text{ext}(\tilde\psi_1)$ is unbounded so we can choose $\textbf{q}\in\text{ext}(\tilde\psi_1)$ with large enough $|\textbf{q}|$ such that $\tilde \psi(\textbf{q})\leq 0$.

Let's assume that a point $\textbf{p}$ exist in $\text{ext}(\tilde\psi_1)$ with $\tilde{\psi}(\textbf{p}) >\tilde \psi_1$. Since $\text{ext}(\psi_1)$ is connected, we can always find a continuous path $\sigma\subseteq\text{ext}(\psi_1)$ from $\textbf{p}$ to $\textbf{q}$. On $\sigma$ path, value of $\tilde\psi$ changes from value $\tilde{\psi}(\textbf{p})$ to a value $\tilde{\psi}(\textbf{q})$.

Now, $\tilde{\psi}(\textbf{p})>\psi_1>0\geq \tilde{\psi}(\textbf{q})$. Hence by intermediate value theorem there exists a point $\textbf{t}$ on $\sigma$ where $\tilde{\psi}(\textbf{t}) = \psi_1$. Hence $\textbf{t}\in\tilde{S}(\tilde{\psi}_1)$. But also, $\textbf{t}\in\sigma\subseteq\text{ext}(\tilde{\psi}_1)$. Given that $\tilde{S}(\tilde{\psi}_1)$,  $\text{ext}(\tilde\psi_1)$ are disjoint, this is a contradiction. Hence, for all of $\text{ext}(\tilde\psi_1)$,  $\tilde{\psi}\leq \tilde\psi_1$.

For any point $\textbf{s}$ on $\tilde{S}(\tilde{\psi}_2)$, $\tilde\psi(\textbf{s})=\tilde\psi_2>\tilde\psi_1$ so they can not exist on $\text{ext}(\psi_1)$. Furthermore, $\psi_2\neq\psi_1$ so they can not exist on $\tilde{S}(\tilde{\psi}_1)$ either. Thus, they must all exist in $\text{int}(\tilde{\psi}_1)$. $\square$\\

\noindent\textbf{Lemma 11} Flux-surfaces with $\tilde{\psi}_- < \tilde{\psi}<\tilde{\psi}_+$ stay strictly in the interior of $\tilde{S}_{\text{in}}$. Flux-surfaces with $0<\tilde{\psi}<\tilde{\psi}_-$ stay strictly in the exterior of $\tilde{S}_{\text{in}}$ and the interior of $\tilde{S}_{\text{out}}$. Here,
\begin{gather}
    \tilde{S}_{\text{in}} = \{\tilde{\textbf{r}}:\tilde\psi(\tilde{\textbf{r}})=\tilde\psi_{\text{in}}\to\tilde{\psi}_-,\ \tilde{\psi}_\text{in}>\tilde{\psi}_-\}
\end{gather}

\noindent\textbf{Proof} We know from \text{lemma 6} that all surfaces except $\tilde{\psi}\in\{0,\psi_-,\psi_+\}$ have no $\nabla\tilde{\psi}=0$ points on them. For our carefully chosen definition of $\tilde{S}_{\text{in}}$, $\tilde{\psi}\to\tilde{\psi}_-$ but it is not equal to $\tilde\psi_-$. Thus, from regular value theorem\cite{GuilleminPollack1974} we can conclude that $\tilde{S}_{\text{in}}$ is a hypersurface. 

For $\tilde{\psi}_-\to\tilde{\psi}_{\text{in}} < \tilde{\psi}$ \textit{lemma 10} gives us that $\tilde{S}(\tilde{\psi})$ is fully in the interior of $\tilde{S}_{\text{in}}=\tilde{S}(\tilde{\psi}_{\text{in}})$. From \textit{lemma 1}, all points with $\tilde{\psi}>0$ are strictly inside the region inside the ellipsoid $\tilde{S}_{\text{out}}$. For $0 <\tilde{\psi} < \tilde{\psi}_-\to{\tilde{\psi}_{\text{in}}}$ \textit{lemma 10} gives us that $\tilde{S}(\tilde{\psi}_{\text{in}})$ is fully in the interior of $\tilde{S}(\tilde{\psi})$, which immediately means $\tilde{S}(\tilde{\psi})$ is fully in the exterior of $\tilde{S}_{\text{in}}$. $\square$\\

\noindent\textbf{Lemma 12} All $\gamma$-fibers with $\tilde{\psi}>0$, or equivalently $\subseteq E\backslash L$, are bounded loops. 

\noindent\textbf{Proof} All flux-surfaces with $\tilde{\psi}>0$ are torus or spherical in topology. $\gamma$ fibers are the intersection of constant $\varphi$ half-planes and flux surfaces for $u>0\implies \psi>0$, so they must intersect in bounded loops, given that flux surfaces are tori or spheres in topology.  $\square$\\

\noindent\textbf{Lemma 13} The domain of $\varphi$ for which a $\gamma$-fibers exists on $\tilde\psi$ is given by,
\begin{gather}
\Theta(\tilde{\psi}) =    \begin{cases}
[0,2\pi)\ \text{if}\ \tilde{\psi}\leq\tilde\psi_-\\
[\varphi_{\text{min}}(\tilde{\psi}),\ 2\pi-\varphi_{\text{min}}(\tilde{\psi})]\ \text{else}
\end{cases}\\
\varphi_{\text{min}}(\tilde{\psi})= \cot^{-1}\left[h_-(\tilde{\psi})/g_-(\tilde{\psi})\right]
\end{gather}
with $g_-$ and $h_-$ defined in Eqs. \eqref{eq: g value} and \eqref{eq: h value}.\\
\noindent\textbf{Proof} This problem implicitly assumes that $u>0$ because $\varphi$ is a well defined concept. We are essentially looking to find the exact form of this set,
\begin{gather}
    \Theta(\tilde{\psi}) = \{\varphi(\tilde{\textbf{r}}):\tilde{\psi}(\tilde{\textbf{r}})=\tilde{\psi}\}
\end{gather}
From lagrange multiplier, we can conclude that an extremum happens whenever $\nabla\varphi\parallel\nabla\tilde{\psi}$ with the constraint that the point is on the $\tilde{\psi}(\tilde{\textbf{r}})=\tilde{\psi}$ surface. We also know that
$\nabla\varphi\parallel\nabla\tilde{\psi}\iff\nabla\tilde{\psi}\times\nabla\varphi = 0$. So, the extremum happens if and only is $\tilde{\textbf{B}} = 0$ on surface $\tilde{\psi}$. Given that $u>0$, only critical points available are $C_{\text{crit}}$. 

The maximum or minimum value of $\varphi$ happens at the intersection of the $C_{\text{crit}}$ circle and flux surfaces. We know from Eqs. \eqref{eq: equiv 1}-\eqref{eq: equiv 4} that there are no intersections if $\psi<\psi_-$. So there are no extremum in the $[0,2\pi)$ range, and the flux surface contains all possible $\varphi$ $\gamma$-fibers. 

For $\tilde{\psi}_-\leq\tilde{\psi}\leq\tilde{\psi}_+$ there is at least one intersection. As before, we focus on $x\geq0$ side WLOG where $\varphi\in[0,\pi]$. We have already calculated the only intersection point on this half to be $(g_-(\tilde{\psi}),h_-(\tilde{\psi}))$ in Eqs. \eqref{eq: g value} and \eqref{eq: h value} during proof of \textit{lemma 8}. At this critical points value of $\varphi$ is $\varphi_{\text{min}}=\cot^{-1}(g_-/h_-)$. Given that $\tilde{x}\geq 0$, $\varphi_{\text{min}}\leq \pi$. We also know from Eq. \eqref{eq: existence of pi} that at least one point exists for $\varphi = \pi$. So clearly $\varphi_{\text{min}}$ is the minimum value of $\varphi$ on surface $\tilde{\psi}$. Given that $x\leq 0$ half of $\tilde{\psi}$ surface is a reflect of $x\geq 0$ half, we deduce that there the $\varphi$ reaches maxima $2\pi-\varphi_{\text{min}}$ instead. So the full range is $[\varphi_{\text{min}}, 2\pi -\varphi_{\text{min}}]$. For $\tilde \psi=\tilde \psi_-$, the intersection happens at $\varphi = 0$ so we assign the range to be $[0,2\pi)$ to ensure no double count. $\square$\\

\noindent\textbf{Lemma 14} The relation $\gamma: D\to \Gamma$, where $\Gamma$ is the set of all field lines, is a bijective function. 
\begin{gather}
    D \equiv \{(\tilde\psi,\varphi):\varphi\in\Theta(\tilde\psi),\ \tilde\psi \leq\tilde\psi_+\} \sqcup \{i\star\}_{i=1}^5
\end{gather}

\noindent\textbf{Proof } Case, a point $p$ exists such that $p\in l\cap L$:  From \textit{lemma 0}, field lines with intersection with $L$ strictly stay on $L$. A field line $l$ will extend in both direction as long as possible as the maximal integral solution to $d\tilde{\textbf{r}}/ds = \tilde{\textbf{B}}$ as long as it does not hit any critical point $\tilde{\textbf{B}} = 0$ Now, there are two 'x' type critical points on $L$ given by $P_{\text{crit}}$. These critical points separate $L$ into five disjoint but connected sections given by $\gamma(i\star)$. The sections $\gamma(2\star)$ and $\gamma(4\star)$ are critical points (and hence field lines) and thus trivially fulfill $l=\gamma$. If $p\in \gamma(1\star), \gamma(3\star)$ or, $\gamma(5\star)$ then $l$ extend until it reaches the limit of the critical points $(0,\tilde{\alpha}m,\tilde{z}_\pm)$. Given that $\gamma(1\star), \gamma(3\star)$ and, $\gamma(5\star)$ are all connected segments, $l$ fills each of them fully in each cases. Thus $l=\gamma(i\star)$ if $p\in \gamma(i\star)$. 

Case, $l\cap L =\varnothing$: Given that $l\cap L =\varnothing$, from
We focus on the case $u>0$ where any $\gamma$ has labeling $(\tilde\psi,\varphi)$. 

For the field lines with a `O' point $\textbf{p}\in C_{\text{crit}}$, the eigenvalues have no real part (see \textit{lemma 2}). Thus, no integral curve approaches or departs from $\textbf{p}$. Hence $\textbf{p}$ itself is the maximal integral curve and thus the associated field line is $l = \{\textbf{p}\}$. From \textit{lemma 0} we know that $l=\{\textbf{p}\}\subseteq \gamma(\tilde{\psi}(\textbf{p}),\varphi(\textbf{p}))$.

We know from \textit{lemma 13} that only surfaces with $\tilde \psi_-\leq\tilde\psi \leq \tilde \psi_+$ intersect $C_{\text{crit}}$ line and whenever they intersect, $\varphi$ is maximized or minimized for that surface. That means the intersection can not contain more than a single point. Hence $\gamma(\tilde{\psi}(\textbf{p}),\varphi(\textbf{p}))$ is a single point, which by definition is $\textbf{p}$. Hence, $l=\gamma(\tilde{\psi}(\textbf{p}),\varphi(\textbf{p})=\{\textbf{p}\}$.

For any point $\textbf{p}\in \gamma(\tilde\psi,\varphi)$, maximally integral of the solution of $d\tilde{\textbf{r}}/ds = \tilde{\textbf{B}}$ with $\textbf{p}$ as an initial point to constructs the field line $l(\textbf{p})$. Given \textit{lemma 0}, $l(\textbf{p})\subseteq\gamma(\tilde\psi,\varphi)$. All the points $\textbf{q}\in l(\textbf{p})$ can equally label the field line as $l(\textbf{q})$, so we instead choose the label $l_\beta$ as the unique label for these field lines. We know that field lines are disjoint by definition, so we can conclude,
\begin{gather}
    \gamma(\tilde\psi,\varphi) = \bigsqcup_{\beta} l_{\beta}
\end{gather}

For $l_\beta$ without any `O' points on them, there exists no `X' critical points on them either due to $u>0$. Thus there are no critical points on $l_\beta$ and they are all closed subsets of $\gamma(\tilde{\psi},\varphi)$. But $\gamma(\tilde\psi,\varphi)$ is a connected set (because it is path connected) and can not be partitioned into more than one non-empty closed subset. Thus $\gamma(\tilde\psi,\varphi)$ consists of a single field line and $l=\gamma(\tilde{\psi},\varphi)$. $\gamma: D\to \Gamma$ is a function.
    
From \textit{lemma 13} and \textit{lemma 7}, $D$ by construction covers all $(\psi,\varphi)$ and $i\star$ for which field lines exist in $\Gamma$. Thus, this is a surjective function. 

If $\gamma(\psi_1,\varphi_1) = \gamma(\psi_2,\varphi_2)$, then for $p\in\gamma(\psi_1,\varphi_1) = \gamma(\psi_2,\varphi_2)$, $\psi(p) =\psi_1,\ \varphi(p)=\varphi_1$ and $\psi(p) =\psi_2,\ \varphi(p)=\varphi_2$ making $(\psi_1,\varphi_1)=(\psi_2,\varphi_2)$. If $\gamma(i_1\star)=\gamma(i_2\star)$, then by disjointness of $\gamma(i\star)$ sets, $i_1\star=i_2\star$. Furthermore, $\gamma(\psi,\varphi) = \gamma(i\star)$ has no solution because  $\gamma(\psi,\varphi)$ requires $\rho>0$ and $\gamma(i\star)$ has $\rho = 0$. So there is no field line that is described by $(\psi,\varphi)$ and $i\star$ labeling at the same time. So, this is an injective function.So $\gamma: D\to \Gamma$ is a bijective function. $\square$

\vspace{-4pt}
\subsection{Finalizing the Proof}
\vspace{-4pt}
All that remains is to collect the lemmas systemically. First, we convert $\tilde\psi_\pm\rightarrow\psi_\pm$ and $\tilde y_\pm\rightarrow y_\pm$ in dimensioned form, which gives us Eq. \eqref{eq: psi_pm}. The dimensioned form for $\tilde{S}_{\text{in}}$ and $\tilde{S}_{\text{out}}$ respectively become $S_{\text{in}}$ and $S_{\text{out}}$. We also switch the "bounded/unbounded lines" convention from set theory to "closed loops/ open lines" to make the theorems fit with physics literature. We now prove the \textit{theorems}.

\begin{enumerate}
\item Immediately follows from \textit{lemma 14}. Furthermore, it proves that field line and $\gamma$-fibers are equivalent objects and thus we can use all lemmas about $\gamma$-fibers interchangeably with field lines. $\blacksquare$ 
    
\item For $\psi\in(-\infty,0)$, the flux-surfaces and field-lines are outside of $S_{\text{out}}$ because of \textit{lemma 1} and open because of \textit{lemma 4}. $\blacksquare$ 
    
\item For $\psi\in(0,\psi_-)$, the flux-surfaces are closed and topologically torus because of \textit{lemma 9}. The flux-surfaces and field-lines are outside $S_{\text{in}}$ but inside $S_{\text{out}}$ because of \textit{lemma 11}. Field-lines are closed loops because of \textit{lemma 12}. $\blacksquare$ 
    
\item For $\psi\in(\psi_-,\psi_+)$, the flux-surfaces are closed and topologically spherical because of \textit{lemma 9}. The flux-surfaces and field-lines are inside $S_{\text{in}}$ because of \textit{lemma 11}. Field-lines are closed because of \textit{lemma 12}. $\blacksquare$ 
\end{enumerate}
\vspace {-4 pt}
\section{Generality of the Perturbation Model}
\vspace{-4pt}
\label{sec: Generality of the Perturbation Model}
The perturbation in Eq. \eqref{eq: perturbation approx} represents a much broader class. 

\textbf{Claim:} Any topological behavior of the vortex under slowly spatially varying, closure-preserving, and zero vorticity perturbations is captured by Eq. \eqref{eq: perturbation approx}.

Zero vorticity in magnetic systems means the absence of current along the field lines, so it can be interpreted as a vacuum condition in the context of the magnetic field. Slowly spatially varying mean perturbations with small wavenumber $k$ such that $kr,kz$ are small in the region of interest, a reasonable assumption for weak perturbations. So, only terms of order $\mathcal{O}{(k r_s+kz_s)}$ will be kept.  Perturbations can be divided into odd- and even-parity types. Weak ($\alpha<0.05$) odd-parity perturbations were conjectured to preserve the closedness of field-lines in the FRC-RMF system in \cite{CohMil}. \cite{Alan} shows that even-parity perturbations open the field-line structure. Thus, the term ``closure preserving" is \textit{defined} to be equivalent to odd-parity in this paper.

\noindent\textbf{Proof of Claim: }We write perturbation in cylindrical co-ordinate $\delta\textbf{B} \equiv (\delta B_r, \delta B_\phi,\delta B_z)$. We defined the Fourier transformation of $\delta \textbf{B}$ in the $(n,k)$ space to be, 
\begin{gather}
   \delta \hat{\textbf{B}}(r,n,k)=\int\int \delta\textbf{B}(r,\phi,z)  e^{-in\phi-ikz} d\phi dz
\end{gather}
As we are requiring the perturbation to vary slowly spatially, this means we will need $\delta \hat{\textbf{B}}(r,n,k)$ with large $k$ to drop off fast. In practice, this means we will only keep $\mathcal{O}(kr)$ and $\mathcal{O}(kz)$ terms in our analysis. In Fourier space, the vacuum condition ($\nabla\times\delta\textbf{B}=0$) gives us, 
\begin{gather}
\frac{in}{r}\delta\hat{B}_z=ik\ \delta\hat{B}_\phi
    \label{eq: B_phi}\\
ik\ \delta\hat{B}_r=\frac{\partial (\delta\hat{B}_z)}{\partial r}
    \label{eq: B_r}\\
   \frac{\partial(r\ \delta\hat{B}_\phi)}{\partial r}=in\delta\hat{B}_r
\end{gather}
And $\nabla\cdot\textbf{B}=0$ gives us,
\begin{gather}
\frac{1}{r}\frac{\partial(r\ \delta\hat{B}_r)}{\partial r}+\frac{in}{r}\ \delta\hat{B}_\phi+ik\ \delta\hat{B}_z=0
\label{eq: div B}
\end{gather}
Replacing $\delta \hat{B}_r$ and $\delta \hat{B}_\phi$ from Eqs. \eqref{eq: B_phi} and \eqref{eq: B_r} in Eq. \eqref{eq: div B} gives us the Bessel equation, 
\begin{gather}
    r^2\frac{\partial^2 (\delta\hat{B}_z)}{\partial r^2}+r\frac{\partial (\delta\hat{B}_z)}{\partial r}-(n^2+k^2r^2)\delta\hat{B}_z=0
    \label{eq: bessel}
\end{gather}
There are two solutions to this: the modified Bessel function of the first kind $I_n(kr)$ and the modified Bessel function of the second kind $K_n(kr)$. $K_n(kr)$ diverges as $kr\rightarrow0$ which is not physical. So, the general solution will only have $\cdot I_n(kr)$ with an arbitrary constant factor. We choose the factor to be $-2\pi\cdot B_0\alpha(n,k)$. The $-2\pi B_0$ constant is added for ease of calculations. This gives us, 
\begin{gather}
    \delta\hat{B}_z= -2\pi B_0 \alpha(n,k) I_n(kr)\\
    \implies  \delta B_z= B_0\int\int \alpha(n,k) I_n(kr) e^{in\phi+kz}\ dn\ dk
\end{gather}

Now, we require all solutions to fulfill $B_z(\phi)=B_z(2\pi+\phi)$. This means $n$ must be an integer. Furthermore, changing the sign of $n$ leaves Eq. \eqref{eq: bessel} unchanged, so the solutions must be identical with the change of sign in $n$. Given that $I_{-n}(kr)=I_n(kr)$, we require $\alpha(n,k)=\alpha(-n,k)\equiv \alpha_n(k)$. The general solution to Eq. \eqref{eq: bessel} in Fourier space will be a superposition of these solutions. After doing a reverse Fourier transform to real space, we thus get, 
\begin{gather}
    \delta B_z= \sum_{n=0}^{\infty} \delta B^{(n)}_z\\
    \delta B^{(0)}_z =-B_0\int \alpha_0(k) I_0(kr) e^{ikz} \ dk\\
    \delta B_z^{(n)}\equiv -2B_0\cos({n\phi)}\int \alpha_n(k) I_n(kr) e^{ikz} \ dk
\end{gather}

In the long wave approximation, we are only interested in terms of order $\mathcal{O}(kr)$ and $\mathcal{O}(kz)$. This means $\alpha(k)$ terms drop off fast for higher $k$ values. $I_n(kr)\sim\mathcal{O}(k^nr^n)$ so we can ignore modes except $n=0$ and $n=1$.

We first focus on $n=0$ terms. Integrating $z$ on both sides of Eqs. \eqref{eq: B_phi} and \eqref{eq: B_r} in real space gives us $\delta B_\phi^{(0)}$ and $\delta B_r^{(0)}$. We require our vector fields to be real. Expanding up to the first order of $k$, we get,
\begin{gather}
\delta B_r^{(0)}= iB_0\int \alpha_0(k) I_1(kr) e^{ikz} dk\approx \frac{B_0 \mu r}{2}\\
\delta B_z^{(0)}= -B_0\int\alpha_0(k)I_0(kr)e^{ikz}dk\approx-B_0 (\nu+2\mu z)
\end{gather}
Where we have defined,
\begin{gather}
    \mu=\frac{i}{2}\int\alpha_0(k)k\ dk,\quad \nu=\int\alpha_0(k) \ dk
\end{gather}
$\delta B^{(0)}_\phi = 0 $ so it can be ignored. We add $\textbf{B}^{(0)}$ to hill's vortex. 
\begin{gather}
    \delta B_r = B_0 \frac{rz}{z_s^2}+B_0 \mu r\\
    \delta B_z = B_0\left(1-\frac{2r^2}{r_s^2}-\frac{z^2}{z_s^2}\right)-B_0 (\nu+2\mu z)  
\end{gather}
We will ignore $\mu^2$ terms as that is of order $k^2$. After simplification, we can show that, 
\begin{gather}
    \delta B_r = B_{0}' \frac{rz'}{z_{s,n}^2},\quad 
    \delta B_z = B_{0}'\left(1- \frac{2r^2}{r_{s}'^2}-\frac{z'^2}{z_{s}'^2}\right)  \\ 
\text{Where,}\quad    z'\equiv z+\mu z_s^2,\quad B_0'\equiv  B_0(1-\nu),\nonumber\\
    r_s'\equiv r_s\sqrt{1-\nu},\quad  z_s'\equiv z_s\sqrt{1-\nu}
\end{gather}
Redefining variables completely absorbs the effects in the hill's vortex terms. The effects are merely scaling and coordinate shifts, so the redefinition keeps the topology unchanged. The impact from the redefinition of constants and coordinate shift in $z$ direction in the $\delta \textbf{B}^{(1)}$ terms can be absorbed by a redefinition of $\alpha_1\rightarrow \alpha_1'=\alpha_1 \exp(-ik\mu z_s^2)/(1-\nu)$. So, $\delta \textbf{B}^{(0)}$ terms have no impact on the topology of the structure and simply rescale the system. So, we will drop the prime superscript and continue to use the variables as they were. 

Now, we focus on the $n=1$ term $\delta \textbf{B}^{(1)}_z$. As before, integrating $z$ on both sides of Eqs. \eqref{eq: B_phi} and \eqref{eq: B_r} in real space gives us $\delta B_\phi^{(1)}$ and $\delta B_r^{(1)}$. We split the solutions into even and odd-parity terms. 
\begin{gather}
    \alpha_1(k)=\alpha_-(k)+i\alpha_+(k)\\
 \text{where,}\   \alpha_\pm(-k)=\pm\alpha_\pm(k)
\end{gather}
This splits the solution into $\delta \textbf{B}^{(1)}=\delta \textbf{B}^{(+)}+\delta \textbf{B}^{(-)}$ terms where the even parity part is,
\begin{gather}
    \delta B^{(+)}_{r}=-2B_0\cos{\phi}\nonumber\\\times\int\alpha_+(k)\left(I_0(kr)-\frac{I_1(kr)}{kr}\right)\cos{(kz)}\ dk\\
    \delta B_\phi^{(+)}=2B_0\sin{\phi}\int \alpha_+(k)\frac{I_1(kr)}{kr}\cos{(kz)}\ dk\\
   \delta B_z^{(+)}=2B_0\cos{\phi}\int\alpha_+(k) I_1(kr)\sin(kz)dk
\end{gather}

The even parity terms completely open up field-lines as proven in \cite{Alan}. We have thus isolated parts of a general perturbation that destroys closure. We require this part to $\alpha_+(k)=0$. What remains is the odd-parity part, which is
\begin{gather}
    \delta B^{(-)}_{r}=-2B_0\cos{\phi}\nonumber\\\times\int\alpha_-(k)  \left(I_0(kr)-\frac{I_1(kr)}{kr}\right)\sin{(kz)}\ dk\\
    \delta B_\phi^{(-)}=2B_0\sin{\phi}\int \alpha_-(k)\frac{I_1(kr)}{kr}\sin{(kz)}\ dk\\
    \delta B_z^{(-)}=-2B_0\cos{\phi}\int\alpha_-(k) I_1(kr)\cos(kz)dk.
\end{gather}

Only the $n=1$ odd-parity perturbations remain in the general perturbation so far, so $\delta\textbf{B} =\delta \textbf{B}^{(-)}$. \textit{It should be noted that this is the static snapshot of the rotating magnetic field used in an FRC.} Now, we expand the odd-parity terms to the first order in $k$ as required by the long wave approximation. 
\begin{gather}
    \delta\textbf{B} =- \langle\alpha\rangle\langle k\rangle  B_0 (z\cos{\phi},-z\sin{\phi},r\cos{\phi})
    \label{eq: final}
\end{gather}
where we have defined,
\begin{gather}
\langle\alpha\rangle\equiv \int \alpha_{-}(k) \ dk,\quad
\langle k\rangle \equiv  \frac{1}{\langle\alpha\rangle}\int \alpha_{-}(k) k\ dk
\end{gather}
Eq. \eqref{eq: final} has effectively the same form as Eq. \eqref{eq: perturbation approx} with $\alpha k$ replaced with $\langle\alpha\rangle\ \langle k\rangle$. So, in summary, the $n=0$ type perturbation only rescales the Hill's vortex, and $n=1$ type closure preserving perturbations can be reduced to the same form as \eqref{eq: perturbation approx}. This proves the claim that our analysis applies to any slow-varying perturbation that preserves closure, which justifies the claim.  $\blacksquare$
\vspace {-4 pt}
\section{Validity of the Vortex Model}
\vspace{-4pt}
\label{sec: Validity of the Vortex Model}
The unperturbed vortex model used in this paper assumed non-zero vorticity (or current in FRC vortices) $\propto r$ everywhere. This is valid within the unperturbed separatrix $r^2/r_s^2+z^2/z_s^2=1$ but not outside. There is nothing mathematically wrong with such a system, but physically it may not be immediately sensible. For example, in the case of FRC, the current drops to zero as one moves away from the FRC core. A more realistic vortex in fluids can thus be modeled as a vortex with non-zero vorticity $\propto r$ inside the unperturbed separatrix and then dropping to $0$ within a layer of thickness $\sigma$ as was done in \cite{orlandi2020instabilities}. In a perturbed vortex, closed field lines, our primary interest, may exist outside of the unperturbed separatrix, where the model has some error. However, if the error is on the order of $\alpha B_0 \cdot\mathcal{O}(k^2r_s^2+k^2z_s^2)$, the topological conclusions from the model remain physically consistent and valid as the error of that order is already ignored. 

\textbf{Claim:} For $0<\alpha\lesssim\alpha_{safe}\equiv \sigma_{min}/r_s\cdot (r_{min}/r_{max})^2$, the topological conclusions about compact flux-surfaces and closed field-lines are valid in our vortex model. They fully break down at $\alpha\sim\alpha_{max}\equiv \alpha_{safe}/(kr_{max})$. Here, $r_{max} \equiv\max{(r_s,z_s)},\ r_{min} \equiv\min{(r_s,z_s)},\  \sigma_{min}\equiv \min(\sigma,r_s,z_s)$.

In real-life vortices, $\sigma/r_s$ is small but non-zero. At the extreme end, $\sigma/r_s$ can go up to $1$ or even higher in FRC vortices, which were studied in \cite{Gota, Steinhauer2020}. Thus, the perturbation can be large without impacting the topology inside the separatrix.

\noindent\textbf{Proof of Claim:} We modeled the unperturbed vortex by Eq. \eqref{eq: field-lines without perturbation}. 

There is an implicit assumption in the model that the current density (or vorticity) is present everywhere. After some calculation, one can show that,
\begin{gather}
     \textbf{J}=\frac{B_0}{\mu_0}\left(\frac{1}{z_s^{2}}+\frac{4}{r_s^2}\right)\cdot r\ \hat{\phi}
     \label{eq: current}
\end{gather}

This means that the current density keeps increasing all the way to infinity, which is not realistic. In a realistic vortex, the current density goes to $0$ outside the separatrix. Thus, outside of the separatrix, the vector fields have a different set of equations than the model we are using. The perturbation we used in Eq. \eqref{eq: perturbation approx} slightly pushes the new outer separatrix by $\Delta r \sim \alpha k (r_s^2+z_s^2)/3$ as can be seen \cite{Ahsan1} and from Eq. \eqref{eq: outerseparatrix}. Roughly within this distance from the unperturbed separatrix $r^2/r_s^2+z^2/z_s^2=1$, the new separatrix does not overlap, and hence, conclusions about closure and topology may not hold. Thus, the vector field has an ambiguous error from using the previous model. If the error is very small compared to the vector field we use in our model, the conclusions in this paper remain valid. 
We parametrize the elliptical separatrix and its immediate neighborhood with the parametric equation,
\begin{gather}
    r=\xi \sin\theta,\quad z = \frac{z_s}{r_s}\xi \cos\theta
\end{gather}
For any ellipse, $\xi$ remains constant and thus can be used as a natural coordinate that respects the boundary condition imposed at $\xi=r_s$ of the ellipse. We also define $\Delta \xi\equiv  \xi-r_s$ as roughly a measure of the distance of an ellipse $\xi$ from the unperturbed separatrix. We now describe a more realistic model for a zero-helicity vortex where the current falls off to $0$ at the edge with thickness $\sigma\ll r_s$, \cite{orlandi2020instabilities}. 
\begin{gather}
    \nabla\times\textbf{B}_0'=\mu_0\textbf{J}',\\
    \textbf{J}'=J_0\sin\theta \cdot \hat{\phi}
    \begin{cases}
        \xi/r_s, &\text{if}\ \xi<r_s\\
        1-\frac{\Delta \xi}{\sigma}&\text{if}\ r_s<\xi<r_s+\sigma\\
        0&\text{if}\ \ r_s+\sigma<\xi\\
    \end{cases}
    \label{eq: model}
\end{gather} 
In the effective model we used in our analysis, we assumed that there is a non-vanishing current proportional to $r$ everywhere. In that model, we have, 
\begin{gather}
    \nabla\times\textbf{B}_0=\mu_0\textbf{J}_0\\
    \textbf{J}_0=J_0\sin\theta \frac{\xi}{r_s}\hat{\phi}= J_0\sin\theta \left(1+\frac{\Delta\xi}{r_s}\right)\hat{\phi}
\end{gather}
Thus, in the $r_s<\xi<r_s+\sigma$ region, we get an error of $\Delta \textbf{B}\equiv \textbf{B}_0-\textbf{B}_0'$. This error fulfills, 
\begin{gather}
    \nabla\times\Delta\textbf{B} = \mu_0J_0\sin\theta \left(\frac{1}{r_s}+\frac{1}{\sigma}\right)\Delta\xi\ \hat{\phi}
\end{gather}
and the boundary condition,
\begin{gather}
\Delta \textbf{B}(r_s,\theta,\phi)=0
\label{eq: BC1}\\
\implies\quad\frac{\partial }{\partial \theta}\Delta \textbf{B}(r_s,\theta,\phi)=\frac{\partial }{\partial \phi}\Delta \textbf{B}(r_s,\theta,\phi)=0
\label{eq: BC2}
\end{gather}
We will express the vector field into components $(\Delta B_\xi,\Delta  B_\theta, \Delta B_\phi)$. The system is symmetric for $\phi$ and hence $\partial_\phi\rightarrow0$. There is no error current in $\xi$ and $\theta$ components. 
\begin{gather}
    (\nabla\times \Delta \textbf{B})_\xi=0,\quad (\nabla\times \Delta \textbf{B})_\theta=0\\\implies \frac{\partial}{\partial\theta}(\Delta B_\phi \sin\theta)=0,\ -\frac{\partial}{\partial \xi}(\xi\Delta B_\phi)=0\\\implies \Delta B_\phi = \frac{C}{\xi\sin\theta}
\end{gather}
From Eq. \eqref{eq: BC1}, $\Delta B_\phi(r_s,\theta,\phi)=0$ which gives us $C=0$. Thus we have 
\begin{gather}
    \Delta B_\phi  = 0, \quad\text{everywhere}
    \label{eq: B phi}\\
    \implies \frac{\partial}{\partial \xi}\Delta B_\phi = 0
    \label{eq: rho partial B phi}
\end{gather}
The $\theta$ component of $\Delta \textbf{B}$ can be analyzed via the curl of the $\phi$ component, and the $\xi$ component can be analyzed via the divergence. At $\xi=r_s$ surface, $\Delta\xi=0$ so, 
\begin{gather}
   \left. \nabla\times\Delta\textbf{B}\right\vert_{\xi=r_s}=\textbf{0}
\end{gather}
$\phi$ component gives us, 
\begin{gather}
    \left.\left(\frac{\partial \Delta B_\theta}{\partial\xi}+\frac{\Delta B_\theta}{\xi}
    -\frac{\partial_\theta \Delta B_\xi}{\xi\sqrt{\sin^2\theta+z_s^2/r_s^2\cdot \cos^2\theta}}\right)\right|_{\xi=r_s}=0
    \label{eq: rho partial B theta 1}
\end{gather}
Given that $\Delta B_\phi=0$ the divergence condition $\nabla\cdot\Delta\textbf{B}=0$ becomes, 
\begin{gather}
    \left.\left(\frac{\partial \Delta B_\xi}{\partial\xi}+\frac{\Delta B_\xi}{\xi}+\frac{\partial_\theta\Delta B_\theta}{\xi\sqrt{\sin^2\theta+z_s^2/r_s^2\cdot\cos^2\theta}}\right)\right|_{\xi=r_s}=0
    \label{eq: rho partial B rho 1}
\end{gather}
Using Eqs. \eqref{eq: BC1} and \eqref{eq: BC2} we can reduce Eqs. \eqref{eq: rho partial B rho 1} and \eqref{eq: rho partial B theta 1} to,
\begin{gather}
    \left.\frac{\partial}{\partial\xi} \Delta B_\theta\right|_{\xi=r_s}=0,\quad \left.\frac{\partial}{\partial\xi} \Delta B_\xi\right|_{\xi=r_s}=0
    \label{eq: rho partial B theta and rho}
\end{gather}   
Summarizing Eqs. \eqref{eq: rho partial B phi} and \eqref{eq: rho partial B theta and rho} we can write, 
\begin{gather}
    \left.\frac{\partial}{\partial \xi}\Delta \textbf{B}\right\vert_{\xi=r_s}=\textbf{0}
    \label{eq: derivative B}
\end{gather}
Given Eqs. \eqref{eq: BC1} and \eqref{eq: derivative B}, after a Taylor expansion up to first order, we can show
\begin{gather}
\Delta \textbf{B}(r_s+\Delta \xi,\theta,\phi)=\textbf{0}+\mathcal{O}(\Delta\xi^2)
\end{gather}
So, the magnitude of error $\Delta B\equiv |\Delta \textbf{B}(r_s+\Delta\xi,\theta,\phi)|$ is $0$ up to first order in $\Delta\xi$. up to the second order, we need to compare it to the dimensioned forms available. The error will be dominated by the smallest length scale $\sigma_{min}$. Thus,
\begin{gather}
    \Delta B\sim \frac{\mu_0J_0\Delta\xi^2}{\sigma_{min}}
\end{gather}
We compare this with the vector field in either model to get an error ratio. We know from Eqs. \eqref{eq: model} and \eqref{eq: current} that current on the separatrix is $J_0$ where,
\begin{gather}
    J_0 = \frac{B_0 r_s}{\mu_0}\left(\frac{1}{z_s^2}+\frac{4}{r_s^2}\right)\sim\frac{B_0 r_s}{\mu_0 r_{min}^2}\\ \implies \Delta B \sim\frac{B_0\Delta\xi^2 r_s}{r_{min}^2\sigma_{min}}
\end{gather}
Ignoring vector fields of order $\alpha B_0 k^2 (r_s^2+z_s^2)\sim \alpha B_0 k^2 r_{max}^2$ is already consistent with our analysis. So the largest $\Delta \xi$, which we label as safe length, that does not have a significant error in our analysis, is,
\begin{gather}
    \frac{B_0\Delta\xi^2 r_s}{r_{min}^2\sigma_{min}}\sim\alpha B_0 k^2 r_{max}^2\\
\implies \Delta\xi_{safe}\sim\sqrt{\frac{\alpha k^2r_{max}^2r_{min}^2\sigma_{min}}{r_s}}
\end{gather}
The theory breaks down when the error is of order equal to the perturbation, which is $\alpha B_0 k (r_s+z_s)\sim \alpha B_0 k\ r_{max}$. 
\begin{gather}
    \frac{B_0\Delta\xi^2 r_s}{r_{min}^2\sigma_{min}}\sim\alpha B_0 k\ r_{max}
\end{gather}
\begin{gather}
    \implies \Delta\xi_{max}\sim\sqrt{\frac{\alpha k \ r_{max}\ r_{min}^2\sigma_{min}}{r_s}}
\end{gather}

The central topological effects, such as closure and compact flux-surfaces, happen inside the new perturbed separatrix. Its center shifts from the unperturbed separatrix by $\alpha k (r_s^2+z_s^2)/3$, and it has a radius of $\sim r_s$. So the largest deviation $\Delta \xi$ we need to worry about in the new separatrix is 
\begin{gather}
    \Delta\xi_{crit}\sim \alpha k\ \frac{r_s^2+z_s^2}{3}\sim\alpha kr_{max}^2
\end{gather}

The topological conclusion remains unaffected when $\Delta\xi_{crit}$ is well within or comparable to the safety region.
\begin{gather}
    \Delta\xi_{crit}\sim\Delta\xi_{safe}
\end{gather}
And the theory will break down when the breakdown error length is comparable to the cutoff length, 
\begin{gather}
    \Delta\xi_{crit}\sim\Delta\xi_{max}
\end{gather}
After straightforward calculations, these conditions reduce to,
\begin{gather}
    \alpha_{safe} \sim \frac{\sigma_{min}}{r_s} \left(\frac{r_{min}}{r_{max}}\right)^2\\
    \alpha_{max} \sim \frac{1}{kr_{max}} \frac{\sigma_{min}}{r_s} \left(\frac{r_{min}}{r_{max}}\right)^2=\frac{\alpha_{safe}}{kr_{max}}.\ \blacksquare
\end{gather}

\bibliographystyle{unsrt}
\bibliography{bibliography}

\end{document}